\documentclass[11pt,a4paper]{article} 
\pdfoutput=1
\usepackage{jheppub,hyperref,mdwlist}
\usepackage{amsmath}
\usepackage{graphicx}
\usepackage{subfigure}
\usepackage{amssymb}
\usepackage{xcolor}
\usepackage{multirow}
\usepackage{cancel}
\usepackage{color}
\usepackage{listings}
\usepackage{xcolor}
\usepackage{float}
\usepackage{slashed}

\usepackage{amsmath}%\ge=\geqslant; \le=\leqslant
\usepackage{wrapfig}

\newcommand{\be}{\begin{equation}}
\newcommand{\ee}{\end{equation}}
\newcommand{\beq}{\begin{equation}}
\newcommand{\eeq}{\end{equation}}
\newcommand{\bea}{\begin{eqnarray}}
\newcommand{\eea}{\end{eqnarray}}
\newcommand{\besp}{\begin{equation}\begin{split}}
\newcommand{\eesp}{\end{split}\end{equation}}

\newcommand{\nn}{\nonumber}

\newcommand{\tr}{\text{tr}}

\newcommand{\Eq}[1]{Eq.~(\ref{#1})}
\newcommand{\Dfbd}{\mathord{\buildrel{\lower3pt\hbox{$\scriptscriptstyle\leftrightarrow$}}\over {D}_{\mu}}}
\newcommand{\ave}[1]{\left\langle #1\right\rangle}
\newcommand{\abs}[1]{\left| #1\right|}

\def\hc{{\rm h.c.}}
\def\mG{\mathcal{G}}
\def\mH{\mathcal{H}}

\def\mL{\mathcal{L}}

\def\mO{\mathcal{O}}

\def\mT{\mathcal{T}}

\def\Z{\mathbb{Z}}

\def\0{\textbf{0}}
\def\1{\textbf{1}}
\def\2{\textbf{2}}
\def\3{\textbf{3}}
\def\4{\textbf{4}}
\def\5{\textbf{5}}
\def\6{\textbf{6}}
\def\7{\textbf{7}}
\def\8{\textbf{8}}
\def\9{\textbf{9}}

\begin{document}

\title{Electroweak baryogenesis and gravitational waves in a composite Higgs model with high dimensional fermion representations}

\author[a]{Ke-Pan Xie,}
\author[1,b]{Ligong Bian,}
\author[1,c]{Yongcheng Wu}
\footnotetext[1]{Corresponding author.}
\affiliation[a]{Center for Theoretical Physics, Department of Physics and Astronomy, Seoul National University, Seoul 08826, Korea}
\affiliation[b]{Department of Physics, Chongqing University, Chongqing 401331, China}
\affiliation[c]{Ottawa-Carleton Institute for Physics, Carleton University, 1125 Colonel By Drive, Ottawa, Ontario K1S 5B6, Canada}

\emailAdd{kpxie@snu.ac.kr}
\emailAdd{lgbycl@cqu.edu.cn}
\emailAdd{ycwu@physics.carleton.ca}

\abstract{We study electroweak baryogenesis in the $SO(6)/SO(5)$ composite Higgs model with the third generation quarks being embedded in the $\2\0'$ representation of $SO(6)$. The scalar sector contains one Higgs doublet and one real singlet, and their potential is given by the Coleman-Weinberg potential evaluated from the form factors of the lightest vector and fermion resonances. We show that the resonance masses at $\mO(1\sim 10{\rm ~TeV})$ can generate a potential that triggers the strong first-order electroweak phase transition (SFOEWPT). The $CP$ violating phase arising from the dimension-6 operator in the top sector is sufficient to yield the observed baryon asymmetry of the universe. The SFOEWPT parameter space is detectable at the future space-based detectors.
}

\maketitle
\flushbottom

\section{Introduction}

The baryon asymmetry of the universe (BAU) is quantitively described by the baryon-to-entropy ratio $\eta_B\equiv n_B/s=[0.82\sim0.94]\times10^{-10}$~\cite{Tanabashi:2018oca}.
The explanation of BAU necessitates the three Sakharov conditions~\cite{Sakharov:1967dj}: i) baryon number non-conservation, ii) $C$ and $CP$ violation, and iii) departure from thermal equilibrium in the early universe. In the Standard Model (SM), although the first condition can be realized via the electroweak (EW) sphaleron~\cite{DOnofrio:2014rug}, the last two conditions are unfortunately not met. The $CP$ violating phase from CKM matrix is too tiny, and the SM EW phase transition (EWPT) is a smooth crossover that cannot provide an out-of-equilibrium environment~\cite{Morrissey:2012db}. Therefore, the observed BAU strongly motivates new physics beyond the SM (BSM). Among various BSM mechanisms accounting for BAU, the EW baryogenesis (EWB) receives extensive attention, especially after the 125 GeV SM-like Higgs boson was discovered at the LHC~\cite{Aad:2012tfa,Chatrchyan:2012ufa}. In the paradigm of EWB, the third Sakharov condition is provided by the strong first-order EWPT (SFOEWPT), and the corresponding BSM physics is typically testable at current or future colliders~\cite{Morrissey:2012db,Arkani-Hamed:2015vfh}. The gravitational waves (GWs) from SFOEWPT are also hopefully detectable at the future space-based detectors~\cite{Mazumdar:2018dfl}.

There have been a lot of researches realizing EWB in the supersymmetric or non-supersymmetric BSM models. As one of the most plausible non-supersymmetric frameworks addressing the SM hierarchy problem, the composite Higgs model (CHM) is an attractive scenario. In this framework, the hierarchy problem is solved by identifying the Higgs doublet as the pseudo-Nambu-Goldstone bosons (pNGBs) from the spontaneous global symmetry breaking $\mG/\mH$ of a new strong interacting sector~\cite{Kaplan:1991dc,Contino:2003ve,Agashe:2004rs}. In CHMs, the SFOEWPT can be triggered by the enlarged scalar sector, either from the dilaton of conformal invariance breaking~\cite{Bruggisser:2018mus,Bruggisser:2018mrt} or from the extra pNGBs of $\mG/\mH$ breaking~\cite{Espinosa:2011eu,Chala:2016ykx,Chala:2018opy,DeCurtis:2019rxl,Bian:2019kmg}; and the new $CP$ phase from the fermion sector can generate BAU~\cite{Bruggisser:2018mus,Bruggisser:2018mrt,Espinosa:2011eu,Chala:2016ykx,Chala:2018opy,DeCurtis:2019rxl}.

In this work we focus on the next-to-minimal CHM (NMCHM), whose coset is $\mG/\mH=SO(6)/SO(5)$, yielding one Higgs doublet plus one real singlet~\cite{Gripaios:2009pe}. It is well-known that such a scalar sector is able to generate a SFOEWPT through the ``two-step'' pattern, providing the essential cosmic environment for EWB~\cite{Cline:2012hg,Alanne:2014bra,Huang:2015bta,Vaskonen:2016yiu,Huang:2017kzu,Huang:2018aja,Chiang:2018gsn,Bian:2018bxr,Bian:2018mkl,Cheng:2018axr,Kurup:2017dzf,Alanne:2019bsm}. However, unlike the normal singlet-extended SM, the NMCHM's scalar potential is generated by the $SO(6)$-breaking terms, which depend on the fermion embeddings in $SO(6)$. As the fermion contribution is dominated by the top quark due to its large mass, hereafter we refer ``fermion embedding'' to the $q_L=(t_L,b_L)^T$ and $t_R$ embeddings. It has been shown that \6 and \1\5 representations are hard to trigger a SFOEWPT, mainly because of the smallness of the quartic couplings~\cite{DeCurtis:2019rxl,Bian:2019kmg}~\footnote{A short comments for other representations lower than \1\5: the \4 gives a large deviation to the $Zb_L\bar b_L$ vertex, while the \1\0 is not able to generate a potential for the singlet thus leaves a massless Goldstone boson in the particle spectrum~\cite{Gripaios:2009pe}. Therefore, they are disfavored by the collider experiments.}. The NMCHM with $q_L$ in \6 and $t_R$ in $\2\0'$ has plenty of parameter space triggering the SFOEWPT since the $\2\0'$ embedding can generate fairly large quartic couplings for the scalars~\cite{DeCurtis:2019rxl}. In this article we consider a NMCHM with $q_L$ and $t_R$ {\it both} in $\2\0'$ (denoted as $\2\0'+\2\0'$). We will demonstrate that a SFOEWPT can be realized by the Coleman-Weinberg potential from the form factors of the lightest composite resonances, and the dimension-6 operator consists of the scalars and top quark provides sufficient $CP$ violation for generating BAU. We also present the study of GW searches for the SFOEWPT parameter space.

This article is organized as follows. The scalar potential generated from the strong sector is studied in Section~\ref{sec:SFOEWPT}, where the possibility of the SFOEWPT is also investigated. The source of $CP$ violation is considered in Section~\ref{sec:EWB}, where we also realize EWB and explain the observed BAU. Section~\ref{sec:GW} is devoted to the GW detectability of the SFOEWPT parameter space. Finally, we conclude in Section~\ref{sec:conclusion}. 
The detailed construction of the $\2\0'+\2\0'$ NMCHM is listed in Appendix~\ref{sec:model}.

\section{The scalar potential and SFOEWPT}\label{sec:SFOEWPT}

\subsection{Sources of the potential}

The NMCHM contains two sectors with different symmetry structures. The elementary sector includes all the SM particles except the Higgs boson, realizing the $SU(2)_L\times U(1)_Y$ gauge symmetry; while the composite sector includes the Higgs and singlet pNGBs and heavier (typically at TeV scale) vector/fermion resonances, experiencing a spontaneous global symmetry breaking $SO(6)/SO(5)$. The interactions between these two sectors preserve the SM gauge group, but break the $SO(6)$ global group explicitly, converting the scalars (i.e. the Higgs and singlet) from exact NGBs to pNGBs, generating the scalar potential and trigger the EW symmetry breaking. In short, interactions between the elementary and composite sectors serve as the sources of the scalar potential.

There are two types of such interactions: fermion mixing (or in the terminology of CHM, ``partial compositeness''~\cite{Agashe:2004rs}) and gauge interaction. Each kind of sources can be further classified into the {\it IR contributions}, coming from the Coleman-Weinberg potential driven by the one-loop form factors of the leading operators of the composite resonances; and the {\it UV contributions}, coming from the local operators generated by the matching of physics at the cut-off scale. The UV contributions are not calculable but only estimated by the na\"ive dimensional analysis (NDA)~\cite{Panico:2011pw}. However, if the UV contributions are negligible, then the potential can be calculated by the Coleman-Weinberg mechanism and expressed as a function of the resonance masses and couplings. This is the so-called {\it minimal Higgs potential hypothesis} (MHP), which is generally adopted in the collider phenomenology studies of the CHMs~\cite{Marzocca:2012zn,Matsedonskyi:2012ym,Pomarol:2012qf,Redi:2012ha,Marzocca:2014msa,Banerjee:2017qod}. In the aspect of cosmological implications, however, NMCHMs with fermion in \1\5 and lower representations cannot give a SFOEWPT under the MHP, due to the small quartic couplings in the potential~\cite{Bian:2019kmg}. In this section, we demonstrate that the $\2\0'+\2\0'$ NMCHM is able to trigger a SFOEWPT, as the quartic coefficients are enhanced in the high-dimensional representation.

In the following two subsections, we will discuss the two sources (partial compositeness and gauge interaction) one by one and derive the potential. The possibility of a SFOEWPT scenario will be investigated in the final subsection.

\subsection{Calculating the scalar potential: fermion contribution}\label{subsec:potential}

The fermion sector contains elementary quarks $q_L=(t_L,b_L)^T$, $t_R$ and composite resonances (also known as top partners, denoted as $\Psi$). According to the partial compositeness mechanism and CCWZ formalism, $q_L$ and $t_R$ should be embedded into the incomplete representations of $SO(6)$ (which we choose as $\2\0'$ in this article), while the top partners are in the complete representations of $SO(5)$ (which we choose as $\1\4$, $\5$ and $\1$ according to the elementary quarks' embedding), and they are connected by the Goldstone matrix $U(h,\eta)$, where $h$ and $\eta$ are the Higgs and singlet, respectively. The details of such construction are listed in Appendix~\ref{sec:model}, where we list all the possible embeddings and select the experimentally favored one, and write down the corresponding Lagrangian. 

In this subsection, we are only interested in the scalar potential at $\mO(100~{\rm GeV})$ scale, hence the heavy top partners can be integrated out, and the relevant degrees of freedom are $q_L$, $t_R$ and the pNGBs $h$, $\eta$. For the sake of deriving the scalar potential, it is just good enough to use the following embeddings
\be\label{top_embedding}
q_L^{\2\0'}=\frac{1}{2}\begin{pmatrix}0_{4\times4} & q_L^\4 & 0_{4\times1} \\
 (q_L^\4)^T & 0 & 0\\
 0_{1\times4}&0&0\end{pmatrix},
 \quad
 t_R^{\2\0'}=\frac{1}{\sqrt{2}}\begin{pmatrix}
 0_{4\times4} & 0_{4\times2} \\
 0_{2\times4} & \sigma^1\,t_R
\end{pmatrix},
\ee
where $q_L^\4\equiv\left(ib_L,b_L,it_L,-t_L\right)^T$, and the Goldstone vector
\be\label{G_vector}
\Sigma=\left(0,0,0,\frac{h}{f},\frac{\eta}{f},\sqrt{1-\frac{h^2+\eta^2}{f^2}}\right)^T,
\ee
to write down the Lagrangian in momentum space up to quadratic level,
\be\begin{split}\label{L_eff}
\mL_{q\Psi}\to&~\tr\left[\bar q_L^{\2\0'}\slashed{p}q_L^{\2\0'}\right]\Pi_0^q+\left(\Sigma^T\bar q_L^{\2\0'}\slashed{p}q_L^{\2\0'}\Sigma\right)\Pi_1^q+\left(\Sigma^T\bar q_L^{\2\0'}\Sigma\right)\slashed{p}\left(\Sigma^Tq_L^{\2\0'}\Sigma\right)\Pi_2^q\\
&+\tr\left[\bar t_R^{\2\0'}\slashed{p}t_R^{\2\0'}\right]\Pi_0^t+\left(\Sigma^T\bar t_R^{\2\0'}\slashed{p}t_R^{\2\0'}\Sigma\right)\Pi_1^t+\left(\Sigma^T\bar t_R^{\2\0'}\Sigma\right)\slashed{p}\left(\Sigma^Tt_R^{\2\0'}\Sigma\right)\Pi_2^t\\
&+\tr\left[\bar q_L^{\2\0'}t_R^{\2\0'}\right]M_0^t+\left(\Sigma^T\bar q_L^{\2\0'}t_R^{\2\0'}\Sigma\right)M_1^t+\left(\Sigma^T\bar q_L^{\2\0'}\Sigma\right)\left(\Sigma^Tt_R^{\2\0'}\Sigma\right)M_2^t+\hc,
\end{split}\ee
where $p$ is the momentum, while $\Pi_{0,1,2}^{q,t}$ and $M_{0,1,2}^{q,t}$ are $p^2$-dependent form factors depending on the strong dynamics.

Substituting Eqs.~(\ref{top_embedding}) and (\ref{G_vector}), \Eq{L_eff} reduces to
\be\label{tLR}
\mL_{q\Psi}\to\left(\bar b_L\slashed{p}b_L\right)\Pi_{LL}^b+\left(\bar t_L\slashed{p}t_L\right)\Pi_{LL}^t+\left(\bar t_R\slashed{p}t_R\right)\Pi_{RR}^t+\left[\left(\bar t_Lt_R\right)\Pi_{LR}^t+\hc\right].
\ee
The form factors for left-handed quarks are
\be\label{Pi_LL}\begin{split}
\Pi_{LL}^b=&~\Pi_0^q+\frac{\Pi_1^q}{2}\frac{\eta^2}{f^2},\quad
\Pi_{LL}^t=\Pi_0^q+\frac{\Pi_1^q}{4}\frac{h^2+2\eta^2}{f^2}+\Pi_2^q\frac{h^2\eta^2}{f^4},\\
\Pi_{LR}^t=&~-\frac{1}{2\sqrt{2}}\frac{h}{f}\sqrt{1-\frac{h^2+\eta^2}{f^2}}\left(M_1^t+4M_2^t\frac{\eta^2}{f^2}\right),
\end{split}\ee
from which we can read the top quark mass
\be
M_t=\frac{1}{2\sqrt{2}}\frac{v}{f}\left|M_1^t\big|_{p^2=0}\right|\sqrt{1-\frac{v^2}{f^2}},
\ee
where $\ave{\eta}=0$ has been used, as required by a SM-like $Zb_L\bar b_L$, see Appendix~\ref{sec:model}.

Given the form factors, the fermion-induced Colman-Weinberg potential is
\be\label{CW}
V_f(h,\eta)=-2N_c\int\frac{d^4Q}{(2\pi)^4}\ln\left(\Pi_{LL}^b\right)-2N_c\int\frac{d^4Q}{(2\pi)^4}\ln\left(\Pi_{LL}^t\Pi_{RR}^t+\frac{|\Pi_{LR}^t|^2}{Q^2}\right),
\ee
where $Q^2=-p^2$ is the Euclidean momentum square, and $N_c=3$ is the SM color number. In the conventional SM where Higgs is an elementary particle, $\Pi_{LL}^{t,b}=1$, $\Pi_{RR}^t=1$ and $\Pi_{LR}^t\propto h$, and hence \Eq{CW} can be integrated analytically, resulting in the well-known top-induced Coleman-Weinberg potential~\cite{Jackiw:1974cv}. However, in NMCHM, the Higgs is a composite pNGB and form factors $\Pi_{LL}^{t,b}$ and $\Pi_{RR,LR}^t$ are $Q^2$-dependent functions via \Eq{Pi_LL}, thus the integral in \Eq{CW} is highly nontrivial. Substituting \Eq{Pi_LL}, one obtains 
\be\label{Vf_original}\begin{split}
V_f(h,\eta)\approx&-2N_c\int\frac{d^4Q}{(2\pi)^4}\left[\ln\left(1+\frac{\Pi_1^q}{2\Pi_0^q}\frac{\eta^2}{ f^2}\right)
+\ln\left(1+\frac{\Pi_1^q}{4\Pi_0^q}\frac{h^2+2\eta^2}{f^2}+\frac{\Pi_2^q}{\Pi_0^q}\frac{h^2\eta^2}{f^4}\right)\right]\\
&-2N_c\int\frac{d^4Q}{(2\pi)^4}\ln\left[1+\frac{\Pi_1^t}{2\Pi_0^t}\left(1-\frac{h^2}{f^2}\right)+\frac{2\Pi_2^t}{\Pi_0^t}\frac{\eta^2}{f^2}\left(1-\frac{h^2+\eta^2}{f^2}\right)\right]\\
&-2N_c\int\frac{d^4Q}{(2\pi)^4}\ln\left[1+\frac{1}{8Q^2\Pi_0^q\Pi_0^t}\frac{h^2}{f^2}\left(1-\frac{h^2+\eta^2}{f^2}\right)\left|M_1^t+4M_2^t\frac{\eta^2}{f^2}\right|^2\right].
\end{split}\ee
Since $f$ is constrained to be $\gtrsim800$ GeV by the current EW and Higgs measurements~\cite{Sanz:2017tco,deBlas:2018tjm}, we expect $\ave{h}^2,\ave{\eta}^2\ll f^2$ at temperatures around and below the EW scale. Therefore, expanding \Eq{Vf_original} to a polynomial of $h^2/f^2$ and $\eta^2/f^2$ is a reasonable approximation. Hence we match \Eq{Vf_original} to
\be\label{V_scalar}
V_f(h,\eta)=\frac{\mu_h^2}{2}h^2+\frac{\mu_\eta^2}{2}\eta^2+\frac{\lambda_h}{4}h^4+\frac{\lambda_\eta}{4}\eta^4+\frac{\lambda_{h\eta}}{2}h^2\eta^2,
\ee
and the coefficients are
\be\label{f_coefficients}\begin{split}
\mu_h^2=&~2\alpha_1^t-\alpha_1^q-\beta_1^tf^2-\frac{1}{2}\epsilon_1f^2,\quad \mu_\eta^2=-8\alpha_2^t-4\alpha_1^q+4\beta_{12}^tf^2,\\
\lambda_h=&~\beta_1^t+\frac14\beta_1^q+\epsilon_1,\quad\lambda_\eta=16\frac{\alpha_2^t}{f^2}+16\beta_2^t+2\beta_1^q-8\beta_{12}^t,\\
\lambda_{h\eta}=&~8\frac{\alpha_2^t}{f^2}-4\frac{\alpha_2^q}{f^2}-8\beta_{12}^t+\frac12\beta_1^q-2\epsilon_{12}+\frac12\epsilon_1,
\end{split}\ee
where the basic integrals are defined as
\be\label{basic_integrals}\begin{split}
&\alpha_{1,2}^{q,t}=\frac{N_c}{f^2}\int\frac{d^4Q}{(2\pi)^4}\frac{\Pi_{1,2}^{q,t}}{\Pi_0^{q,t}};\\
&\beta_{1,2}^{q,t}=\frac{N_c}{f^4}\int\frac{d^4Q}{(2\pi)^4}\left(\frac{\Pi_{1,2}^{q,t}}{\Pi_0^{q,t}}\right)^2,\quad
\beta_{12}^{q,t}=\frac{N_c}{f^4}\int\frac{d^4Q}{(2\pi)^4}\frac{\Pi_{1}^{q,t}\Pi_{2}^{q,t}}{(\Pi_0^{q,t})^2};\\
&\epsilon_{1,2}=\frac{N_c}{f^4}\int\frac{d^4Q}{(2\pi)^4}\frac{\abs{M_{1,2}^t}^2}{Q^2\Pi_0^q\Pi_0^t},\quad
\epsilon_{12}=\frac{N_c}{f^4}\int\frac{d^4Q}{(2\pi)^4}\frac{\left(M_{1}^t\right)^*M_{2}^t+\left(M_{2}^t\right)^*M_{1}^t}{Q^2\Pi_0^q\Pi_0^t}.
\end{split}\ee
Now the coefficients in \Eq{V_scalar} are all expressed in terms of the momentum integrals of the form factors. Compared to the corresponding equations in the NMCHM with fermion embeddings in \1\5 or \6~\cite{Bian:2019kmg}, the $\lambda_{\eta,h\eta}$ here receive the leading order contribution from the $\alpha$ integrals and thus are enhanced~\footnote{A large quartic coupling in the Coleman-Weinberg potential can also be realized by triplet-singlet mixings~\cite{Csaki:2019coc}.}.

For a QCD-like theory, the form factors can be explicitly written as a sum of the resonance poles
\be\label{Pif}\begin{split}
\Pi_0^{q,t}=&~1+\sum_{n=1}^{N_{\1\4}}\frac{|y_{L,R}^{\1\4(n)}|^2f^2}{Q^2+M_{\1\4(n)}^2},\quad
\Pi_1^{q,t}=2\left(\sum_{n=1}^{N_{\5}}\frac{|y_{L,R}^{\5(n)}|^2f^2}{Q^2+M_{\5(n)}^2}-\sum_{n=1}^{N_{\1\4}}\frac{|y_{L,R}^{\1\4(n)}|^2f^2}{Q^2+M_{\1\4(n)}^2}\right),\\
\Pi_2^{q,t}=&~\frac65\sum_{n=1}^{N_{\1}}\frac{|y_{L,R}^{\1(n)}|^2f^2}{Q^2+M_{\1(n)}^2}-2\sum_{n=1}^{N_{\5}}\frac{|y_{L,R}^{\5(n)}|^2f^2}{Q^2+M_{\5(n)}^2}+\frac45\sum_{n=1}^{N_{\1\4}}\frac{|y_{L,R}^{\1\4(n)}|^2f^2}{Q^2+M_{\1\4(n)}^2},\\
\end{split}\ee
and
\bea
M_0^{t}&=&\sum_{n=1}^{N_{\1\4}}\frac{y_L^{\1\4(n)}y_R^{\1\4(n)*}f^2M_{\1\4(n)}}{Q^2+M_{\1\4(n)}^2},\nn\\
M_1^{t}&=&2\left(\sum_{n=1}^{N_{\5}}\frac{y_L^{\5(n)}y_R^{\5(n)*}f^2M_{\5(n)}}{Q^2+M_{\5(n)}^2}-\sum_{n=1}^{N_{\1\4}}\frac{y_L^{\1\4(n)}y_R^{\1\4(n)*}f^2M_{\1\4(n)}}{Q^2+M_{\1\4(n)}^2}\right),\\
M_2^t&=&\frac65\sum_{n=1}^{N_{\1}}\frac{y_L^{\1(n)}y_R^{\1(n)*}f^2M_{\1(n)}}{Q^2+M_{\1(n)}^2}-2\sum_{n=1}^{N_{\5}}\frac{y_L^{\5(n)}y_R^{\5(n)*}f^2M_{\5(n)}}{Q^2+M_{\5(n)}^2}+\frac45\sum_{n=1}^{N_{\1\4}}\frac{y_L^{\1\4(n)}y_R^{\1\4(n)*}f^2M_{\1\4(n)}}{Q^2+M_{\1\4(n)}^2},\nn
\eea
where $N_{\1\4}$ is the number of top partners in \1\4 representation of $SO(5)$, denoted as $\Psi_{\1\4(n)}$ with $n=1,2,\cdots,N_{\1\4}$. For the $n$-th $\Psi_{\1\4}$, its mass and mixing couplings with $q_L$, $t_R$ are denoted as $M_{\1\4(n)}$ and $y_{L,R}^{\1\4(n)}$, respectively. Similar notations are for $\Psi_{\5(n)}$ and $\Psi_{\1(n)}$. In general, at large $Q^2$ the form factors scale as $\Pi_{1,2}^{q,t},M_{1,2}^t\sim Q^{-2}$. That means that in \Eq{basic_integrals} the $\epsilon$ integrals are convergent, while the $\alpha$ and $\beta$ integrals diverge quadratically and logarithmically, respectively. Inspired by the successful experience in QCD~\cite{Contino:2010rs,Weinberg:1967kj,Shifman:1978bx,Shifman:1978by,Knecht:1997ts}, people apply the Weinberg sum rules to the form factor integrals in CHMs to get a finite scalar potential~\cite{Bian:2019kmg,Marzocca:2012zn,Pomarol:2012qf,Redi:2012ha,Marzocca:2014msa,Banerjee:2017qod}~\footnote{See Ref.~\cite{Marzocca:2012zn} for a detailed discussion on the Coleman-Weinberg potential and the Weinberg sum rules in the CHMs.  In terms of fermions, the Weinberg sum rules can be implemented in terms of a new symmetry, i.e., the maximal symmetry~\cite{Csaki:2017cep,Csaki:2018zzf}. 
}. We will also adopt this assumption here. The convergence of the $\alpha$ and $\beta$ integrals requires $\Pi_{1,2}^{q,t}\sim Q^{-6}$, which can be achieved by imposing the following sum rules
\be\label{SR_3}\begin{split}
\sum_{n=1}^{N_{\1\4}}|y_{L,R}^{\1\4(n)}|^2=&~\sum_{n=1}^{N_{\5}}|y_{L,R}^{\5(n)}|^2=\sum_{n=1}^{N_{\1}}|y_{L,R}^{\1(n)}|^2,\\
\sum_{n=1}^{N_{\1\4}}|y_{L,R}^{\1\4(n)}|^2M_{\1\4(n)}^2=&~\sum_{n=1}^{N_{\5}}|y_{L,R}^{\5(n)}|^2M_{\5(n)}^2=\sum_{n=1}^{N_{\1}}|y_{L,R}^{\1(n)}|^2M_{\1(n)}^2.
\end{split}\ee

Once \Eq{SR_3} is satisfied, the coefficients $\mu_{h,\eta}^2$ and $\lambda_{h,\eta,h\eta}$ in \Eq{V_scalar} are finite functions of the top partner masses and mixing couplings. However, the concrete features of the coefficients depend on the particle content we choose, i.e. depend on $(N_{\1\4},N_\5,N_\1)$. For example, under the simplest setup $(1,1,1)$, \Eq{SR_3} implies equal masses and mixing parameters for all the top partners, thus $\Pi_{1,2}^{q,t}\equiv0$ and hence $\alpha_{1,2}^{q,t}=0$ and $\beta_{1,2,12}^{q,t}=0$, which, after substituting into \Eq{f_coefficients}, gives $\ave{h}=\sqrt{-\mu_h^2/\lambda_h}=f/\sqrt{2}$. This is obviously inconsistent with the EW measurement. The next-to-minimal $(N_{\1\4},N_\5,N_\1)$ contents are also ruled out based on the following considerations: the $(1,1,2)$ gives $\lambda_h=0$ thus EWSB cannot be triggered; the $(1,2,1)$ implies $\lambda_\eta<0$ thus the potential is not bounded below; the $(2,1,1)$ is very likely to have $\mu_\eta^2>0$ and the necessary condition of SFOEWPT is not satisfied. Finally, we find the next-to-next-to-minimal setup $(N_{\1\4},N_\5,N_\1)=(2,1,2)$ has the potential to trigger a SFOEWPT. In this case, the sum rules reduce to
\be\label{SR212}\begin{split}
|y_{L,R}^{\1\4}|^2+|y_{L,R}^{\1\4'}|^2=&~|y_{L,R}^{\5}|^2=|y_{L,R}^{\1}|^2+|y_{L,R}^{\1'}|^2,\\
|y_{L,R}^{\1\4}|^2M_{\1\4}^2+|y_{L,R}^{\1\4'}|^2M_{\1\4'}^2=&~|y_{L,R}^{\5}|^2M_{\5}^2=|y_{L,R}^{\1}|^2M_{\1}^2+|y_{L,R}^{\1'}|^2M_{\1'}^2,
\end{split}\ee
where we denote the two top partners in \1\4 as $\Psi_{\1\4}$ and $\Psi_{\1\4'}$, with the latter being the heavier one. Similar notation also applies to $\Psi_\5$ and $\Psi_{\5'}$. \Eq{SR212} implies
\be
M_{\1\4}<M_{\5}<M_{\1\4'},\quad M_{\1}<M_{\5}<M_{\1'}.
\ee
For the form factors we have
\be\begin{split}
\Pi_0^{q,t}=&~1+\frac{|y_{L,R}^{\1\4}|^2f^2}{Q^2+M_{\1\4}^2}+\frac{|y_{L,R}^{\1\4'}|^2f^2}{Q^2+M_{\1\4'}^2},\quad
\Pi_1^{q,t}=-\frac{2|y_{L,R}^{\1\4'}|^2f^2(M_{\1\4'}^2-M_{\1\4}^2)(M_{\1\4'}^2-M_\5^2)}{(Q^2+M_{\1\4}^2)(Q^2+M_{\1\4'}^2)(Q^2+M_{\5}^2)},\\
\Pi_2^{q,t}=&~\frac{4}{5}\frac{|y_{L,R}^{\1\4'}|^2f^2(M_{\1\4'}^2-M_{\1\4}^2)(M_{\1\4'}^2-M_\5^2)}{(Q^2+M_{\1\4}^2)(Q^2+M_{\1\4'}^2)(Q^2+M_{\5}^2)}+\frac{6}{5}\frac{|y_{L,R}^{\1'}|^2f^2(M_{\1'}^2-M_{\1}^2)(M_{\1'}^2-M_\5^2)}{(Q^2+M_{\1}^2)(Q^2+M_{\1'}^2)(Q^2+M_{\5}^2)}.
\end{split}\ee
Substituting above expressions to \Eq{basic_integrals} and then \Eq{V_scalar}, the fermion-induced potential is now a function of the top partner masses and couplings. 

\subsection{Calculating the scalar potential: gauge contribution}

The vector sector includes elementary gauge bosons $W_\mu^a$ and $B_\mu$ and the composite resonances. We consider the spin-1 resonances in \1\0 and \5 representations of $SO(5)$, and denote them as $\rho_\mu$ and $a_\mu$ respectively. Again, the details of the Lagrangian are given in Appendix~\ref{sec:model} and here we only focus on the content relevant to scalar potential generation. Since only a subgroup $SU(2)_L\times U(1)_Y$ is gauged, the whole $SO(6)$ group is explicitly broken down to $SU(2)_L\times U(1)_Y\times U(1)_\eta$~\cite{Gripaios:2009pe}, with $U(1)_\eta$ being the subgroup generated by the transform along the $\eta$ direction. Therefore, we can expect that gauge interactions only generate potential for $h$, not for $\eta$.

The effective Lagrangian of the vector sector after integrating out the resonances are
\be\label{EW_form}
\mL_\rho\to\frac12P_T^{\mu\nu}\left(-p^2B_\mu B_\nu-p^2\tr\left[W_\mu W_\nu\right]+\Pi_0\tr\left[A_\mu A_\nu\right]+\Pi_1\Sigma^\dagger A_\mu A_\nu\Sigma\right),
\ee
where $\Pi_{0,1}$ are $p^2$-dependent form factors, and $gA_\mu=gW_\mu^aT_L^a+g' B_\mu T_R^3$. The transverse projection operator is $P_T^{\mu\nu}=g^{\mu\nu}-(p^\mu p^\nu)/p^2$. Expanding in components, \Eq{EW_form} reduces to
\begin{multline}\label{H_form_g}
\mL_\rho\to\frac12P_T^{\mu\nu}\left\{\left(-p^2+\frac{g'^2}{g^2}\Pi_0\right)B_\mu B_\nu+\left(-p^2+\Pi_0\right)W_\mu^a W_\nu^a\right.\\
\left.+\frac{\Pi_1}{4}\frac{h^2}{f^2}\left[W_\mu^1W_\nu^1+W_\mu^2W_\nu^2+\left(W_\mu^3-\frac{g'}{g}B_\mu\right)\left(W_\nu^3-\frac{g'}{g}B_\nu\right)\right]\right\},
\end{multline}
from which we can read the Higgs potential as~\cite{Agashe:2004rs}
\be\label{V_g}
V_g(h)\approx\frac{3}{2}\int\frac{d^4Q}{(2\pi)^4}\left\{2\ln\left(1+\frac{\Pi_1}{4\Pi_W}\frac{h^2}{f^2}\right)
+\ln\left[1+\left(\frac{g'^2}{g^2}\frac{\Pi_1}{4\Pi_B}+\frac{\Pi_1}{4\Pi_W}\right)\frac{h^2}{f^2}\right]\right\},
\ee
where $Q^2\equiv-p^2$ and $\Pi_W=Q^2+\Pi_0$, $\Pi_B=Q^2+(g'^2/g^2)\Pi_0$. No potential for $\eta$ is generated, as expected. In the conventional SM, $\Pi_{W,B}=Q^2$ and $\Pi_1=g^2f^2$ and \Eq{V_g} is just the known $W$-induced Coleman-Weinberg potential. However, in NMCHM, the form factors $\Pi_{0,1}$ have nontrivial dependence on $Q^2$ and \Eq{V_g} is affected by the strong dynamics. Expanding \Eq{V_g} up to $h^4$ level gives a very good approximation since the higher order terms are suppressed by $g^2h^2/f^2$. Hence we can write
\be\label{V_gg}
V_g(h)\approx\frac{\mu_g^2}{2}h^2+\frac{\lambda_g}{4}h^4,
\ee
where
\be\label{g_contribution}\begin{split}
\mu_g^2=&~\frac{3}{4f^2}\int\frac{d^4Q}{(2\pi)^4}\left(\frac{g'^2}{g^2}\frac{\Pi_1}{\Pi_B}+3\frac{\Pi_1}{\Pi_W}\right),\\
\lambda_g=&~-\frac{3}{16f^4}\int\frac{d^4Q}{(2\pi)^4}\left[2\left(\frac{\Pi_1}{\Pi_W^2}\right)^2+\left(\frac{g'^2}{g^2}\frac{\Pi_1}{\Pi_B}+\frac{\Pi_1}{\Pi_W}\right)^2\right].
\end{split}\ee

Similar to the fermion-induced case, the form factors $\Pi_{0,1}$ are the sum of the vector resonance poles~\cite{Contino:2010rs}
\be\label{Pi_definition}\begin{split}
\Pi_0=&~g^2Q^2\sum_{n=1}^{N_\rho}\frac{f_{\rho(n)}^2}{Q^2+M_{\rho(n)}^2},\\
\Pi_1=&~g^2f^2+2g^2Q^2\left(\sum_{n=1}^{N_a}\frac{f_{a(n)}^2}{Q^2+M_{a(n)}^2}-\sum_{n=1}^{N_\rho}\frac{f_{\rho(n)}^2}{Q^2+M_{\rho(n)}^2}\right),
\end{split}\ee
where $N_\rho$ is the number of the vector resonances in \1\0 of $SO(5)$. The mass and coupling for $n$-th $\rho_\mu$ is denoted as $M_{\rho(n)}$ and $f_{\rho(n)}$, respectively, satisfying $f_{\rho(n)}\equiv M_{\rho(n)}/g_{\rho(n)}$. Similar notations are used for the $a_\mu$ resonances. To get a convergent $\mu_g^2$ and $\lambda_g$, we impose the Weinberg first and second sum rules
\be
\sum_{n=1}^{N_\rho}f_{\rho(n)}^2=\frac{f^2}{2}+\sum_{n=1}^{N_a}f_{a(n)}^2;\quad
\sum_{n=1}^{N_\rho}f_{\rho(n)}^2M_{\rho(n)}^2=\sum_{n=1}^{N_a}f_{a(n)}^2M_{a(n)}^2,
\ee
so that the scaling of $\Pi_1$ is changed to $Q^{-4}$. Assuming the lightest resonances dominate, i.e. $N_\rho=N_a=1$, the rules reduce to
\be\label{sum_rules_I}
f_\rho^2=\frac{f^2}{2}+f_a^2,\quad f_\rho^2M_\rho^2=f_a^2M_a^2,
\ee
which give
\be
\Pi_0=g^2Q^2\frac{f_{\rho}^2}{Q^2+M_{\rho}^2},\quad
\Pi_1=\frac{g^2f^2M_\rho^2M_a^2}{(Q^2+M_\rho^2)(Q^2+M_a^2)},
\ee
and then the integral in \Eq{g_contribution} can be evaluated analytically~\cite{Marzocca:2012zn,Bian:2019kmg}
\be\label{g_IR}\begin{split}
\mu_g^2=&~\frac{3(3g^2+g'^2)}{64\pi^2}\frac{M_\rho^2M_a^2}{M_a^2-M_\rho^2}\ln\frac{M_a^2}{M_\rho^2},\\
\lambda_g=&~\frac{3\left[2g^4+(g^2+g'^2)^2\right]}{256\pi^2(M_a^2-M_\rho^2)^2}\left[M_a^4+\frac{M_\rho^4(M_\rho^2-3M_a^2)}{M_a^2-M_\rho^2}\ln\frac{M_a^2}{M_W^2}+(a\leftrightarrow\rho)\right].
\end{split}\ee

\subsection{SFOEWPT}\label{subsec:SFOEWPT}

Summing up $V_f(h,\eta)$ in \Eq{V_scalar} and $V_g(h)$ in \Eq{V_gg}, we get the total scalar potential $V(h,\eta)$ of the $\2\0'+\2\0'$ NMCHM. We will still use the coefficient notation in \Eq{V_scalar}, with the definitions of $\mu_h^2$ and $\lambda_h$ absorbing the gauge contributions. At zero temperature, the vacuum of $V(h,\eta)$ is $(\ave{h},\ave{\eta})=(v,0)$, where $v=\sqrt{-\mu_h^2/\lambda_h}$. The field shift for a physical Higgs boson is
\be
h\to v+\sqrt{1-\frac{v^2}{f^2}}h,
\ee
where the factor involving $f$ is due to the higher order operators in the Goldstone kinetic term, i.e. \Eq{NGB_kinetic}. The potential is shifted to
\be\begin{split}
V\to&~-\mu_h^2\left(1-\frac{v^2}{f^2}\right)h^2+\lambda_hv\left(1-\frac{v^2}{f^2}\right)^{3/2}h^3+\frac{\lambda_h}{4}\left(1-\frac{v^2}{f^2}\right)^2h^4\\
&~+\frac{1}{2}(\mu_\eta^2+\lambda_{h\eta}v^2)\eta^2+\frac{\lambda_\eta}{4}\eta^4+\lambda_{h\eta}v\sqrt{1-\frac{v^2}{f^2}}h\eta^2+\frac{\lambda_{h\eta}}{2}\left(1-\frac{v^2}{f^2}\right)h^2\eta^2,
\end{split}\ee
from which we can read the tree-level physical masses of the scalars
\be\label{scalar_mass}
M_h^2=-2\mu_h^2\left(1-\frac{v^2}{f^2}\right),\quad M_\eta^2=\mu_\eta^2+\lambda_{h\eta}v^2.
\ee
Since $v^2\ll f^2$, the observed $M_h=125.09$ GeV and $v=246$ GeV almost fix $\mu_h^2$ and $\lambda_h$. The scalar interacting vertices are also obtained easily.

At the LHC, the singlet $\eta$ can be produced by $gg$ fusion via the SM quarks/top partners triangle loop, or from the decay of Higgs or composite resonances (e.g. $\rho_D$, $\Psi_\1$, etc). The possible decay channels of $\eta$ are model-dependent, including the SM di-boson (induced by the WZW anomaly~\cite{Gripaios:2009pe}) and di-jet (gluon or quark). The $\eta$ can even be a dark matter candidate if it has an odd $\Z_2$ quantum number~\cite{Frigerio:2012uc,Marzocca:2014msa,Ma:2017vzm,Cacciapaglia:2018avr}. Note that although our potential $V(h,\eta)$ and the third generation fermion couplings are both symmetric under $\eta\to-\eta$, a $\Z_2$-breaking term can generally arise from the WZW anomaly or the fermion embeddings of quarks in the first two generations or leptons. As long as $M_\eta>M_h/2$ so that the Higgs exotic decay $h\to\eta\eta$ is kinematically forbidden, the direct search bounds on $\eta$ are not very strong~\footnote{Even for $M_\eta<M_h/2$, there is still rooms for SFOEWPT without conflict with current data~\cite{Kozaczuk:2019pet,Carena:2019une}.}. A scalar of $M_\eta\sim\mO(100~{\rm GeV})$ is still allowed~\cite{Cacciapaglia:2019zmj,Belyaev:2016ftv,Cacciapaglia:2017iws,Cacciapaglia:2019bqz}.

Thermal corrections to the potential can be derived using the finite temperature field theory. Since the vector and fermion resonances are at the $\mO({\rm TeV})$ scale, at temperature around EW scale they can be integrated out and we only deal with the SM degrees of freedom plus the singlet $\eta$. 
%{\color{red}
Therefore, the thermal potential is~\footnote{The contribution from the higher order derivative terms of \Eq{NGB_kinetic} is at most percent level and hence can be safely dropped~\cite{DeCurtis:2019rxl}.}
\be\label{VT_scalar}
V_T(h,\eta)=V(h,\eta)\pm\sum_j \frac{n_jT^4}{2\pi^2}\int_0^\infty x^2dx\ln\left[1\mp e^{-\sqrt{x^2+M_j^2(h,\phi)/T^2}}\right]+V_{\rm daisy}(h,\eta)\;,
\ee
The factor $n_j$ represents number of degree of freedom for scalars, vector bosons, and top quarks. The field-dependent masses $M_j(h,\eta)$ are
\be\begin{split}
&M_h^2=\mu_h^2+3\lambda_hh^2+\lambda_{h\eta}\eta^2,\quad
M_\eta^2=\mu_\eta^2+3\lambda_\eta\eta^2+\lambda_{h\eta}h^2,\quad
M_{h\eta}^2=2\lambda_{h\eta}h\eta,\\
&M_{G^\pm}^2=M_{G^0}^2=\mu_h^2+\lambda_h h^2+\lambda_{h\eta}\eta^2,
\end{split}\ee
for the scalars (where $G^{\pm,0}$ denote the Goldstone modes of the Higgs doublet) and
\be
M_W^2=\frac{g^2}{4}h^2,\quad M_Z^2=\frac{g^2+g'^2}{4}h^2,\quad M_t^2=\frac{y_t^2}{2}h^2.
\ee
for the vector bosons and fermion (in which we only consider top quark due to its sizable mass). Here, $g$, $g^{\prime}$, and $y_t$ are the EW gauge couplings and top Yukawa, respectively. $V_{\rm daisy}$ is the daisy resummation correction for the scalars and longitudinal mode of the vector bosons, i.e.
\be
V_{\rm daisy}(h,\eta)=-\sum_{j'}\frac{T}{12\pi}\left[M_{j'}^3(h,\eta,T)-M_{j'}^3(h,\eta)\right],
\ee
where $j'$ runs over the bosons, and the expressions for thermal mass $M_{j'}^3(h,\eta,T)$ can be found in Refs.~\cite{Carrington:1991hz,Espinosa:2011ax}.

The thermal potential in \Eq{VT_scalar} can trigger a so-called two-step cosmic phase transition, in which the VEV changes as
\be
(h=0,\eta=0)\to(h=0,\eta\neq0)\to(h\neq0,\eta\sim0),
\ee
as the universe cools down. The first-step is a second-order phase transition along the $\eta$ direction, while the second-step is a first-order EWPT via the VEV flipping between the $\eta$- and $h$-axises.
%}
The onset of the first-order EWPT occurs at the nucleation temperature $T_n$ defined by
\be
T_n^4e^{-S_3(T_n)/T_n}\approx H^4(T_n),
\ee
where $S_3$ is the Euclidean action of the $O(3)$ bounce solution~\cite{Linde:1981zj}, and $H(T)$ is the Hubble constant. Numerically, for $T_n\sim100$ GeV the above condition reduces to~\cite{Quiros:1999jp}
\be\label{n_condition}
\frac{S_3(T_n)}{T_n}\sim140.
\ee
If the Higgs VEV at $T_n$ further satisfies
\be\label{S_condition}v_n/T_n\gtrsim1,
\ee
then the EW sphaleron process is suppressed inside the bubble~\cite{Moore:1998swa}, and hence the generated baryon number is not washed out. This is essential for EWB. A first-order EWPT satisfying \Eq{S_condition} is called a SFOEWPT.

As Section~\ref{subsec:potential} has expressed $V(h,\eta)$ as a function of the resonance masses and couplings, realizing SFOEWPT in the $\2\0'+\2\0'$ NMCHM is just to find the parameter space that generates a $V(h,\eta)$ satisfying Eqs.~(\ref{n_condition}) and (\ref{S_condition}). Numerically, we use the following parameters as inputs
\be\label{inputs}
\left\{f, M_{\1\4}, M_{\1\4'}, M_\5, M_\1, M_{\1'},  y_L^\5 \right\},
\ee
and evaluate $y_R^\5$, $y_{L,R}^{\1\4}$, $y_{L,R}^{\1\4'}$, $y_{L,R}^{\1}$ and $y_{L,R}^{\1'}$ (all treated as real numbers in this section) by the Weinberg sum rules \Eq{SR212} and the requirement of top mass $M_t=150$ GeV (the running mass at TeV scale~\cite{Sirunyan:2019jyn}). Then the fermion-induced potential is calculated by performing the $Q^2$ integral for the form factors in \Eq{f_coefficients}. The gauge-induced part, which is determined by $M_\rho$ and $M_a$ in \Eq{g_IR}, is derived by requiring the Higgs and $W$ boson masses to be the experimentally measured ones, i.e. $M_h=125$ GeV, $M_W=80.4$ GeV~\cite{Tanabashi:2018oca}. By this procedure, given a set of parameters in \Eq{inputs}, one gets a scalar potential $V(h,\eta)$ reproducing the SM particle mass spectrum. After that, we use the {\tt MultiNest} package~\cite{Feroz:2008xx} combining with the {\tt CosmoTransitions}~\cite{Wainwright:2011kj} package to calculate $S_3$ and check whether the SFOEWPT is triggered.

\begin{figure}
\centering
\subfigure{
\includegraphics[scale=0.415]{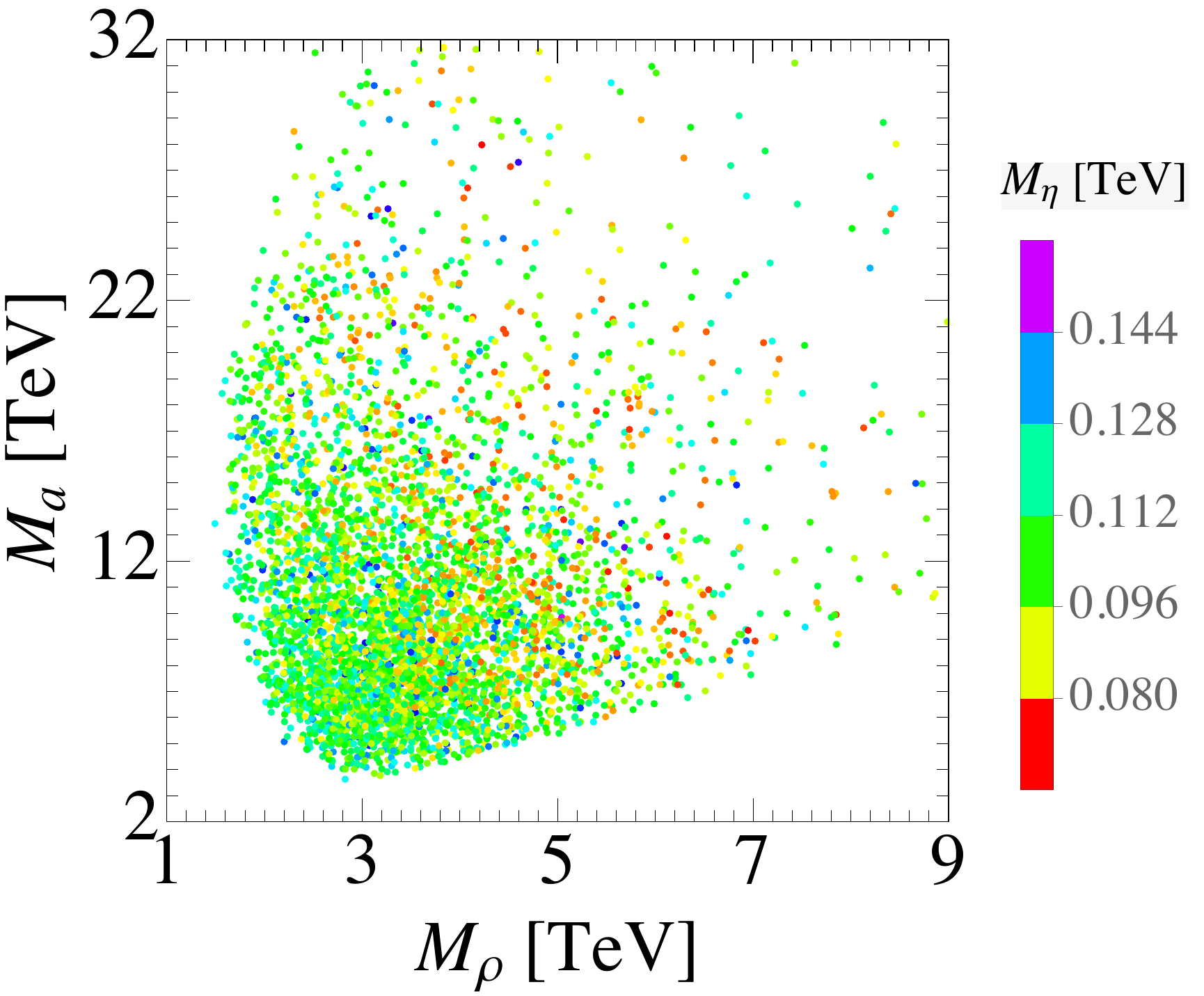}}\quad
\subfigure{
\includegraphics[scale=0.4]{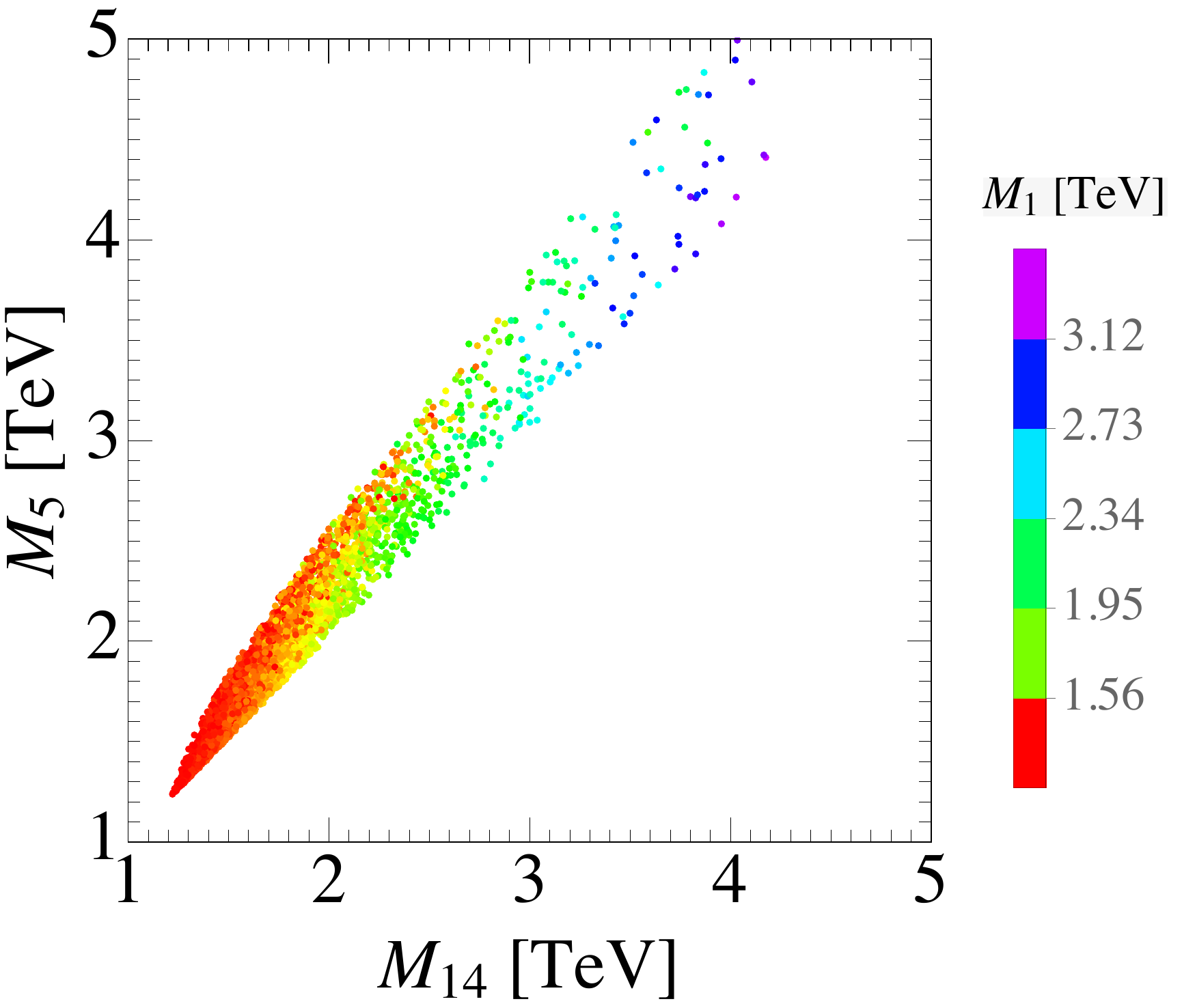}}
\caption{Projecting the parameter points with successful SFOEWPT into the mass planes of boson resonances ($\rho$, $a$) and top partners ($\Psi_{\1\4}$, $\Psi_\5$). The masses of $\eta$ and $\Psi_\1$ are shown in color.}
\label{fig:SFOEWPT}
\end{figure}

As shown in Fig.~\ref{fig:SFOEWPT}, SFOEWPT can be achieved by $M_\eta\sim\mO(100~{\rm GeV})$, $M_{\rho,a}\sim\mO(1\sim10~{\rm TeV})$ and $M_{\1\4,\5,\1}\sim\mO({\rm TeV})$. The magnitudes of the mixing parameters $y_{L,R}$ are smaller than 5, while $1<g_\rho<4\pi$. We have also checked that including the higher order expansions (e.g. $h^6$, $\eta^6$, etc) in the Coleman-Weinberg potential only gives $\lesssim2\%$ corrections to the VEVs at $T_c$ or $T_n$. This confirms the validity of our treatment that keeps only the terms up to quartic-level. At the EW scale, the Lagrangian of NMCHM can be matched to an effective field theory (EFT) formalism with the SM particles, heavy vector multiplets and vector-like quarks as ingredients~\footnote{The mixing of fermions after imposing the $\ave{\eta}=0$ condition is especially simple
\be\begin{split}
\mL_{q\Psi}\to\mL_{\rm EFT}\supset&~\left(y_R^{\1\4}s_Lc_R-\frac{y_L^\5}{\sqrt2}c_Ls_R\right)\bar q_L\widetilde Ht_R+\left(\frac{y_L^\5}{\sqrt2}s_Ls_R+y_R^{\1\4}c_Lc_R\right)\bar J_{QL}\widetilde Ht_R\\
&~-\left(\frac{y_L^\5}{\sqrt2}c_Lc_R+y_R^{\1\4}s_Ls_R\right)\bar q_L\widetilde H\widetilde T_R-y_R^{\1\4}c_R\bar J_{XL}Ht_R,
\end{split}\ee
where
\be
s_R=\frac{y_R^\5f/\sqrt2}{\sqrt{M_\5^2+(y_R^\5 f)^2/2}},\quad s_L=\frac{y_L^{\1\4}f}{\sqrt{M_{\1\4}^2+(y_L^{\1\4}f)^2}},\quad c_{L,R},\equiv\sqrt{1-s_{L,R}^2}
\ee
and $\widetilde T$, $J_{Q}$ and $J_X$ are top partners decomposed from the $\Psi_{\1,\1\4}$ multiplets, see Appendix~\ref{sec:model} for the details.}. We check the indirect constraints from the oblique parameters, which has been measured to be $S=0.02\pm0.07$ and $T=0.06\pm0.06$~\cite{Tanabashi:2018oca}. The contributions from Higgs, spin-1 and spin-1/2 resonances can be found in Refs.~\cite{Contino:2010rs},~\cite{Ghosh:2015wiz} and~\cite{Panico:2010is} respectively. Only the points successfully pass the EW precision test (i.e. not excluded by the oblique parameter bounds at 95\% C.L.) are shown in Fig.~\ref{fig:SFOEWPT}, in which the mixing angles between the top quark and top partners are $\lesssim0.08$.

\section{Electroweak baryogenesis}\label{sec:EWB}

Previous sections have demonstrated that the $\2\0'+\2\0'$ NMCHM can trigger the SFOEWPT for a large range of parameter space. In this section, we study the $CP$ non-conservation sources and calculate the BAU. In Section~\ref{subsec:SFOEWPT}, while deriving the parameter space for SFOEWPT we treated the couplings (e.g. $y_{L,R}^{\1\4}$) as real numbers. However, in general they can be complex. Omitting the complex phases in the fermion couplings is valid for the SFOEWPT study because $CP$ violation only has a minor impact on the phase transition dynamics. But in the study of BAU, those phases are crucial. In \Eq{Yukawa} there are $2(N_{\1\4}+N_\5+N_\1)$ complex phases in the $y_{L,R}^{\1\4,\5,\1}$ couplings, while $(N_{\1\4}+N_\5+N_\1+1)$ of them can be absorbed by the fermion fields, remaining $(N_{\1\4}+N_\5+N_\1-1)$ physical ones. For our chosen particle content $(N_{\1\4},N_\5,N_\1)=(2,1,2)$, there are 4 physical $CP$ violating phases.

At the EW scale, after integrating out the top partners, the $CP$ phases manifest themselves as the complex Wilson coefficients of the operators,
\be\label{top_CPV}
\mL_{q\Psi}\supset-\frac{1}{2\sqrt{2}}\bar t_Lt_R\frac{h}{f}\left(M_{1,0}^t+4M_{2,0}^t\frac{\eta^2}{f^2}\right)\sqrt{1-\frac{h^2+\eta^2}{f^2}}+\hc,
\ee
where
\be\begin{split}
M_{1,0}^{t}\equiv&~ M_1^t\big|_{Q^2=0}=2f^2\left(\frac{y_L^{\5}y_R^{\5*}}{M_{\5}}-\frac{y_L^{\1\4}y_R^{\1\4*}}{M_{\1\4}}-\frac{y_L^{\1\4'}y_R^{\1\4'*}}{M_{\1\4'}}\right),\\
M_{2,0}^{t}\equiv&~ M_2^t\big|_{Q^2=0}=-M_{1,0}^t
+\frac65f^2\left(\frac{y_L^{\1}y_R^{\1*}}{M_{\1}}+\frac{y_L^{\1'}y_R^{\1'*}}{M_{\1'}}-\frac{y_L^{\1\4}y_R^{\1\4*}}{M_{\1\4}}-\frac{y_L^{\1\4'}y_R^{\1\4'*}}{M_{\1\4'}}\right),
\end{split}\ee
are complex numbers. For later convenience, we parametrize \Eq{top_CPV} as
\be\label{top_CPV2}\begin{split}
\text{\Eq{top_CPV}}\approx&-\frac{y_t}{\sqrt{2}}\bar t_Lt_Rh\left[\frac{M_{1,0}^t}{|M_{1,0}^t|}\left(1-\frac{h^2-v^2}{2f^2}\right)+\frac{\eta^2}{2f^2}\left(\frac{8M_{2,0}^t}{|M_{1,0}^t|}-\frac{M_{1,0}^t}{|M_{1,0}^t|}\right)\right]+\hc\\
\equiv&-\frac{y_t}{\sqrt{2}}\bar t_Lt_Rh\left[e^{i\phi_1}\left(1-\frac{h^2+\eta^2-v^2}{2f^2}\right)+\rho_t e^{i\phi_2}\frac{\eta^2}{2f^2}\right]+\hc
\end{split}\ee
where $y_t=\sqrt{2}M_t/v$ is the top Yukawa coupling, and $\rho_t$ and $\phi_{1,2}$ are real numbers derived from the $y_{L,R}$ coefficients. The phase $\phi_1$ can always be absorbed by the redefinition of $t_R$, while $\phi_2$ is the physical phase that characterizes the magnitude of $CP$ violation. In this scenario, the $CP$ non-conservation comes from the dimension-6 operator $ih\eta^2\bar t\gamma^5t$ where the constraints from the electric dipole moment (EDM) measurements are weak due to the absence of mixing between 
$h$ and $\eta$ at tree or loop level. This is different from the dimension-5 operator $ih\eta\bar t\gamma^5 t$ in previous studies~\cite{Espinosa:2011eu,Chala:2016ykx,Chala:2018opy,DeCurtis:2019rxl} where the mixing between $h$ and $\eta$ arises after integrating out the top quark, and then the $CP$ phase suffers from server constraints from 
EDM measurements~\cite{Morrissey:2012db}, especially the measure of electron EDM by ACME~\cite{Andreev:2018ayy}~\footnote{The study of the EWB with SM EFT is also confronting tension with the EDM experimental measurements, see Refs.~\cite{deVries:2017ncy,Balazs:2016yvi,Ellis:2019flb}.}.

During the SFOEWPT, $h$ and $\eta$ are treated as spacetime-dependent background fields. In the rest frame of the bubble wall, the profiles of the scalars (denoted as $\hat h$ and $\hat\eta$) depend only on $z$ and have a kink shape with a wall width $L_{\rm w}$. Near the wall one can treat the profile as a one dimensional problem with the coordinate origin being stabilized at the wall center, and the $z$ axis perpendicular to the wall.

For the two-step phase transition scenario we consider, the bubble wall is usually ``thick'' in the sense that $L_{\rm w}\gtrsim p_z^{-1}$, where $p_z\sim T_n$ is the typical magnitude of the $z$-component momentum of particles in the thermal bath. For example, the numerical results in Ref.~\cite{Konstandin:2014zta} show that $L_{\rm w}\gtrsim10/T_n$. The $CP$ violating interactions nearby the bubble wall create a chiral asymmetry, which is then converted into a baryon asymmetry via the EW sphaleron process, and swept into the bubble when the wall passes by. 
 Inside the bubble, the sphaleron process is frozen by $v_n/T_n\gtrsim1$, thus the baryon asymmetry survives, yielding the observed BAU~\cite{Morrissey:2012db}. This is the non-local EWB mechanism proposed by Refs.~\cite{Joyce:1994fu,Joyce:1994zt}, and we will apply it to the $\2\0'+\2\0'$ NMCHM case in this work~\footnote{For a recent study on local EWB we refer to Ref.~\cite{Zhou:2020xqi}.}.

Technically, we adopt the framework of Ref.~\cite{Fromme:2006wx} to calculate the BAU~\footnote{The framework of Ref.~\cite{Fromme:2006wx} only applies to the subsonic $v_{\rm w}$, while recently a new study~\cite{Cline:2020jre} provides a novel treatment valid for the whole range of $v_{\rm w}\in[0,1]$.}. First, we substitute the bounce solutions and rewrite \Eq{top_CPV2} to the following ``complex mass'' form
\be\label{complex_mass}
\mL_{q\Psi}\supset-m_t\,\bar t\, e^{i\gamma^5\theta_t}t,
\ee
where $m_t$ and $\theta_t$ are $z$-dependent functions determined by
\be\label{complex_mass_2}
m_t=\frac{y_t}{\sqrt{2}}\hat h\left[1-\frac{\hat h^2-v^2}{2f^2}-\frac{\hat\eta^2}{2f^2}\left(1-\rho_t\cos\phi_2\right)\right],\quad
\tan\theta_t=\frac{\hat\eta^2}{2f^2}\rho_t\sin\phi_2.
\ee
The excess of $t_L$ against $t_R$ is calculated by a set of coupled Boltzmann equations, see Ref.~\cite{Fromme:2006wx,Jiang:2015cwa}. The BAU is generated by integrating over the region in the EW unbroken phase~\cite{Fromme:2006wx,Cline:2000nw}
\be
\eta_B=\frac{n_B}{s}=\frac{405\Gamma_{\rm ws}}{4\pi^2v_{\rm w}g_*T_n}\int_0^\infty dz\mu_{B_L}(z)e^{-\frac{45\Gamma_{\rm ws}}{4v_{\rm w}}z},
\ee
where $\Gamma_{\rm ws}\approx18\,\alpha_W^5T_n$ is the EW sphaleron rate outside the bubble~\cite{DOnofrio:2014rug}, $g_*\sim100$ is the number of relativistic degrees of freedom at $T_n$, $\mu_{B_L}(z)$ is the chemical potential of the left-handed quarks (all three generations), and $v_{\rm w}$ is the bubble expansion velocity relative to the plasma {\it just in front of} the bubble wall. Due to the lack of a detailed simulation of the hydrodynamics in the plasma, we use $v_{\rm w}=0.1$ and $L_{\rm w}=15/T_n$ as a benchmark.

\begin{table}
\scriptsize\centering
% {\color{red}
\begin{tabular}{|c|c|c|c|c|c|c|c|c|c|}\hline
 & $f$ [TeV] & $M_\rho$ [TeV] & $M_a$ [TeV] & $M_{\1\4}$ [TeV] & $M_\5$ [TeV] & $M_{\1}$ [TeV] & $M_{\1\4'}$ [TeV] & $M_{\1'}$ [TeV] \\ \hline
B1 & 1.61 & 2.20 & 10.7 & 1.47 & 1.65 & 1.08 & 7.78 & 11.3 \\ \hline
B2 & 1.92 & 3.14 & 8.16 & 1.55 & 1.81 & 1.05 & 7.88 & 12.3 \\ \hline
\end{tabular}
\begin{tabular}{|c|c|c|c|c|c|c|c|c|c|c|c|c|c|c|}\hline
& $y_L^{\1\4}$ & $y_R^{\1\4}$ & $y_L^\5$ & $y_R^\5$ & $y_L^{\1}$ & $y_R^{\1}$ & $y_L^{\1\4'}$ & $y_R^{\1\4'}$ & $y_L^{\1'}$ & $y_R^{\1'}$ & $M_\eta$ [GeV] \\ \hline
B1 & 1.67 & 0.641 & $-1.68$ & 0.642 & 1.67 & 0.638 & 0.166 & 0.0635 & 0.186 & 0.0713 & 108 \\ \hline
B2 & 1.77 & 0.658 & 1.78 & $-0.663$ & 1.77 & 0.658 & 0.216 & 0.0804 & 0.215 & 0.0800 & 92.9 \\ \hline
\end{tabular}
\begin{tabular}{|c|c|c|c|c|c|c|c|c|c|c|c|c|c|c|}\hline
& $\mu_h^2$ [GeV$^2$] & $\mu_\eta^2$ [GeV$^2$] & $\lambda_h$ & $\lambda_\eta$ & $\lambda_{h\eta}$ \\ \hline
B1 & $-(89.5)^2$ & $-(89.4)^2$ & 0.132 & 0.332 & 0.324 \\ \hline
B2 & $-(89.2)^2$ & $-(96.9)^2$ & 0.131 & 0.357 & 0.297 \\ \hline
\end{tabular}
% }
\caption{The benchmarks used to evaluate the BAU. The $T_n$ for B1 and B2 are respectively 104 GeV and 88.8 GeV; while $v_n$ for B1 and B2 are respectively 210 GeV and 225 GeV. The coefficients of the potential are also shown, where $\mu_h^2$ and $\lambda_h$ are almost fixed by the SM Higgs mass and VEV.}\label{tab:EWB_benchmarks}
\end{table}

\begin{figure}
\centering
\subfigure{
\includegraphics[scale=0.418]{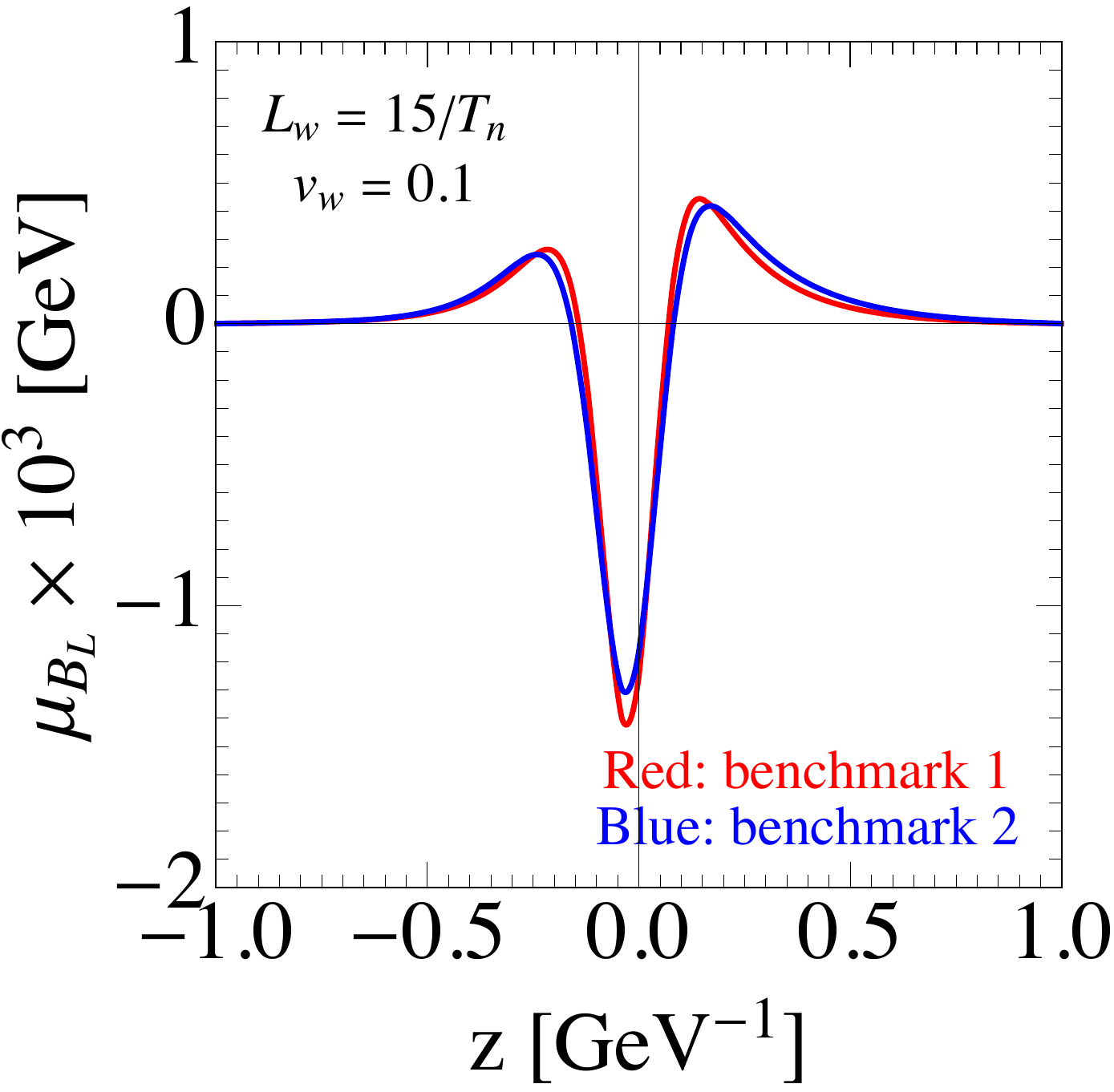}}\qquad
\subfigure{
\includegraphics[scale=0.4]{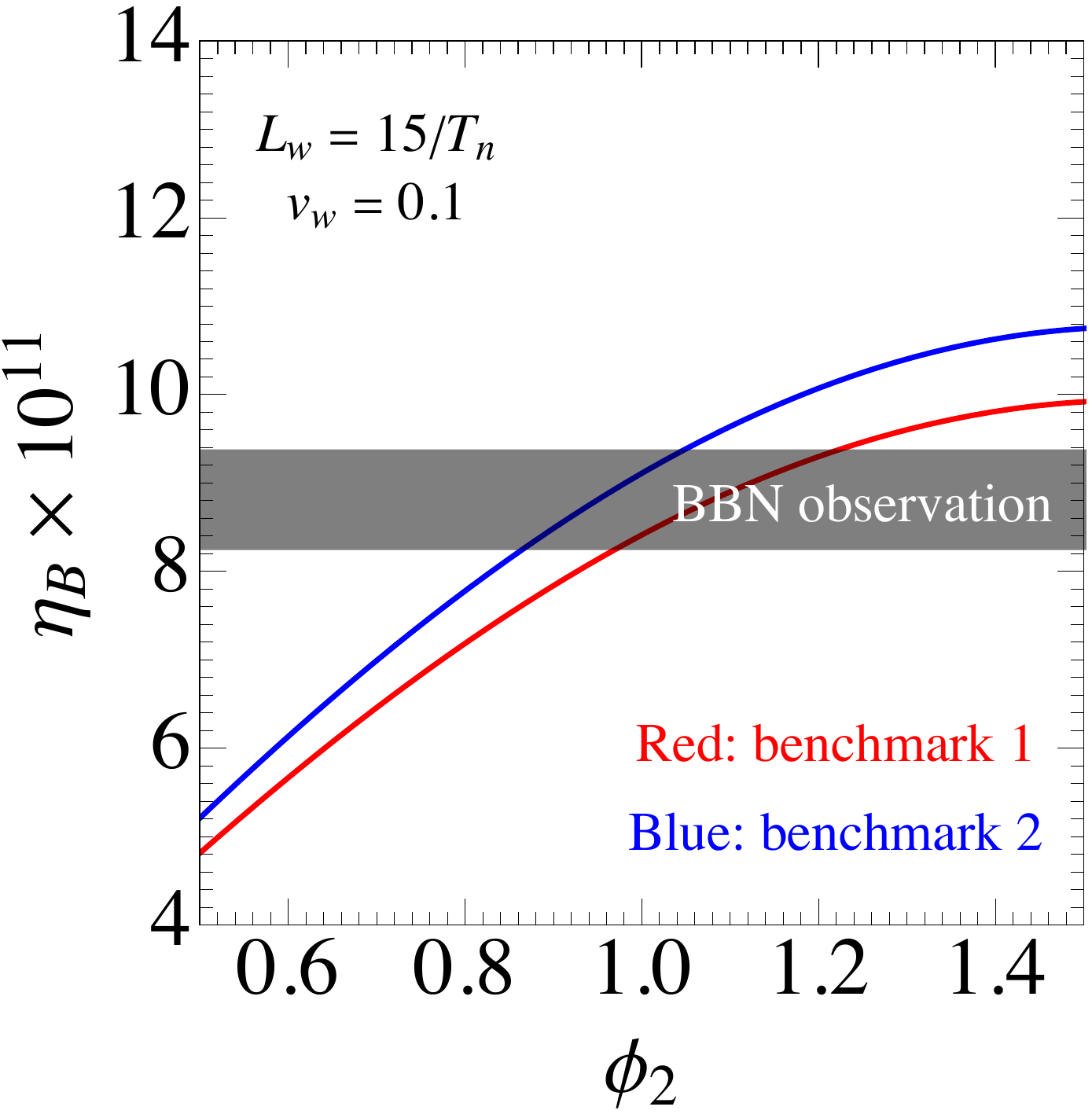}}
\caption{The $\mu_{B_L}(z)$ profiles (solved from $\phi_2=\pi/2$) and BAU from benchmarks B1 and B2. The gray band in the right panel stands for the observed BAU from the Big Bang nucleosynthesis~\cite{Tanabashi:2018oca}.}
\label{fig:baryogenesis}
\end{figure}

Given the bubble profiles and the $CP$ phase $\phi_2$, $\eta_B$ is evaluated straight forward using the equations in Ref.~\cite{Fromme:2006wx}. We confirm that the observed BAU can be reached using the SFOEWPT parameter points derived in Section~\ref{subsec:SFOEWPT}. To illustrate this, we select two benchmarks as listed in Table~\ref{tab:EWB_benchmarks}. The resolved chemical potentials of the benchmarks are plotted in the left panel of Fig.~\ref{fig:baryogenesis}, while the generated BAU are plotted in the right panel as functions of the $\phi_2$. We see that the observed BAU can be explained in the two benchmarks.

\section{Gravitational waves}\label{sec:GW}

An important consequence of the SFOEWPT is the stochastic GWs. For a SFOEWPT that happens at $T_n\sim100$ GeV, the frequency of the GW signal peak is typically mille-Hz after the cosmological redshift~\cite{Grojean:2006bp}, within the sensitive signal region of a set of near-future space-based GW detectors, such as LISA~\cite{Audley:2017drz} and its possible successor BBO~\cite{Crowder:2005nr}, TianQin~\cite{Luo:2015ght,Hu:2017yoc}, Taiji~\cite{Hu:2017mde} or DECIGO~\cite{Kawamura:2011zz,Kawamura:2006up}. The phase transition GWs result from three sources, i.e. collision of the vacuum bubbles, sound waves in the fluid, and the turbulence in plasma. The spectrum of the GWs is described by
\be
\Omega_{\rm GW}(f)=\frac{1}{\rho_c}\frac{d\rho_{\rm GW}}{d\ln f},
\ee
where $\rho_c$ is the critical energy density in the present universe. For the GWs induced by the first-order cosmic phase transition, the spectra can be written in numerical functions of three parameters~\cite{Grojean:2006bp,Caprini:2015zlo}:
\begin{enumerate}
\item $\alpha$, the ratio of EWPT latent heat to the energy density of the universe at $T_n$:
\be
\alpha=\frac{\epsilon}{\rho_{\rm rad}},\quad\epsilon=-\Delta V_T+T_n\Delta\frac{\partial V_T}{\partial T}\Big|_{T_n},\quad  \rho_{\rm rad}=\frac{\pi^2}{30}g_*T_n^4,
\ee
here ``$\Delta$'' denotes the difference between the true and false vacua. Larger $\alpha$ produces stronger GWs.

\item $\beta/H_n$, where $\beta^{-1}$ is the time duration of the EWPT, while $H_n$ is the Hubble constant at $T_n$, i.e.
\be
\beta=\frac{d}{dt}\left(\frac{S_3}{T}\right)\Big|_{t=t_n},\quad
\frac{\beta}{H_n}=T_n\frac{d}{dT}\left(\frac{S_3}{T}\right)\Big|_{T=T_n},
\ee
with $t_n$ being the cosmic time at $T_n$. The smaller $\beta/H_n$ is, the longer EWPT lasts and the stronger GWs are produced.

\item $\tilde v_{\rm w}$, defined as the wall velocity with respect to the plasma at {\it infinite distance}. Note that $\tilde v_{\rm w}$ can be significantly different from $v_{\rm w}$~\cite{No:2011fi}, which is the relative wall velocity to plasma {\it in front of the wall} (defined in Section~\ref{sec:EWB}). $v_{\rm w}$ is relevant for baryogenesis, while $\tilde v_{\rm w}$ is important in the GWs strength calculation. We adopt $\tilde v_{\rm w}=0.6$ as a benchmark.
\end{enumerate}

Using the numerical results in Ref.~\cite{Caprini:2015zlo}, we can express the GW signal strengths in terms of $\alpha$, $\beta/H_n$ and $\tilde v_{\rm w}$. For the benchmarks we consider, the dominant source of the GWs is the sound waves~\cite{Caprini:2015zlo}~\footnote{The detailed studies on the sound waves from a SFOEWPT can be found in Refs.~\cite{Ellis:2018mja,Schmitz:2020rag}.}.
The nucleation temperature $T_n$ is shown in color. To investigate the sensitivity of LISA to the GWs, we evaluate the signal-to-noise ratio (SNR) defined as follows~\cite{Caprini:2015zlo}
\be
{\rm SNR}=\sqrt{\mT\int_{f_{\rm min}}^{f_{\rm max}} df\left(\frac{\Omega_{\rm GW}(f)}{\Omega_{\rm LISA}(f)}\right)^2},
\ee
where $\Omega_{\rm LISA}$ is the sensitive curve of the LISA detector~\cite{Audley:2017drz}, and $\mT$ is the data-taking duration, which is taken to be $75\%\times4$ years, i.e. $9.46 \times 10^7$ s~\cite{Caprini:2019egz}. 

\begin{figure}
\centering
\subfigure{
\includegraphics[scale=0.4]{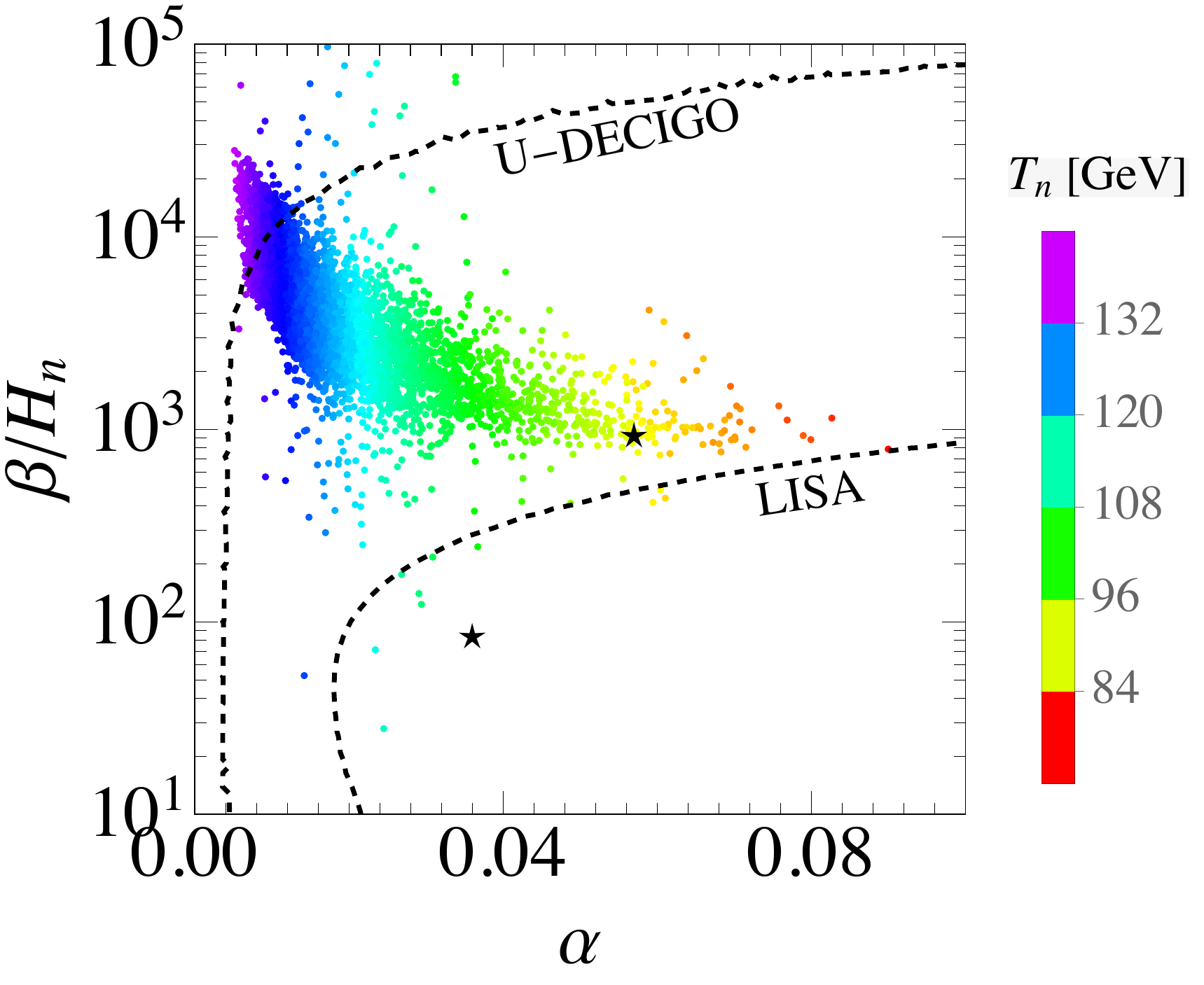}} 
\caption{Distributions of $\alpha$, $\beta/H_n$ and $T_n$ for parameter points with SFOEWPT. {The ${\rm SNR}=10$ for LISA and the U-DECIGO reach is shown as dashed curves.} The EWB benchmarks in Section~\ref{sec:EWB} are highlighted as stars.}
\label{fig:GWs}
\end{figure}

% {\color{red}
We calculate $\alpha$ and $\beta/H_n$ for each parameter points with SFOEWPT, and show the results in Fig.~\ref{fig:GWs}.  Following Ref.~\cite{Caprini:2015zlo}, we adopt ${\rm SNR}=10$ as the detection threshold of LISA. For the U-DECIGO detector, due to the lack of a detailed SNR study, we simply assume that a GW signal is detectable if its peak strength exceeds the sensitivity curve of U-DECIGO. TianQin and Taiji may provide a search complementary to LISA, and we leave the quantitive study of those two detectors to a future work.
% }

\section{Conclusion}\label{sec:conclusion}

In this paper, we studied EWB in the $SO(6)/SO(5)$ CHM, i.e. the NMCHM. The scalar sector contains one Higgs doublet $H$ and one real scalar $\eta$, and the concrete form of potential depends on the fermion embeddings in $SO(6)$. In this work we considered the third generation quarks $q_L=(t_L,b_L)$ and $t_R$ both in the $\2\0'$. According to the decomposition of $SO(6)\times U(1)_X\to SU(2)_L\times U(1)_Y$, there are three and two ways to embed $q_L$ and $t_R$, respectively. To protect the $Zb_L\bar b_L$ vertex, the specific embedding $q_L^{\2\0'_A}$ and $t_R^{\2\0'_B}$ is chosen, and used as the $\2\0'+\2\0'$ NMCHM for the cosmological study.

The scalar potential $V(h,\eta)$ is derived using the one-loop Coleman-Weinberg potential of the form factors from the lightest resonances $\rho$, $a$ and $\Psi_{\1\4,\5,\1}$. Making use of the Weinberg sum rules, the form factor integrals are convergent and a finite $V(h,\eta)$ is evaluated as a function of the resonance masses and couplings. With the help of numerical tools, we found a lot of parameter points that give the SM particle spectrum and the SFOEWPT. The real singlet mass is $\mO(100~{\rm GeV})$, while the vector and fermion resonance masses are typically $\mO(1\sim10{\rm ~TeV})$, thus they are hopefully probed at the LHC. To our best knowledge, this is the first composite Higgs model that succeeds to trigger the SFOEWPT completely via the Coleman-Weinberg potential contributed from the resonances. At the EW scale, the new $CP$ violating phase $\phi_2$ arises from the complex Wilson coefficient of a dimension-6 operator $ih\eta^2\bar t\gamma^5t$ in the top sector. The observed BAU can be explained by suitable value of $\phi_2$ using the non-local EWB mechanism. Also, a considerable fraction of the SFOEWPT points give detectable GW signals at the near-future detectors.

\acknowledgments

We are grateful to Jing Shu and Katsuya Hashino for the useful discussions. We thank Jian-Dong Zhang for communication on the TianQin project. We also thank the anonymous referee for useful suggestions. LGB is supported by the National Natural Science Foundation of China under grant No.12075041, No.11605016 and No.11947406. YCW is supported by the Natural Sciences and Engineering Research Council of Canada (NSERC). KPX is supported by the Grant Korea NRF 2015R1A4A1042542 and NRF 2017R1D1A1B03030820.

\appendix

\section{The \texorpdfstring{$\2\0'+\2\0'$}{20'+20'} NMCHM}\label{sec:model}

Below the confinement scale of the CHMs, the relevant physical degrees of freedom are the pNGBs and the composite resonances, and the effective Lagrangian can be written using the Coleman-Callan-Wess-Zumino (CCWZ) formalism~\cite{Coleman:1969sm,Callan:1969sn}. In this appendix, we only quote the main results in the first two subsections, while the full expressions of the formulae can be found in the final subsection~\footnote{For a nice introduction to the application of CCWZ in the CHMs, we refer the readers to Ref.~\cite{Panico:2015jxa}. See Refs.~\cite{Li:2019ghf,Qi:2019ocx} for the effective field theory studies on CHMs.}.

\subsection{The scalar and vector sectors}

Symmetry breaking pattern is the crucial part of the CCWZ construction. For the NMCHM, the $SO(6)$ group contains 15 generators, which can be chosen as $T^A=\{T^{\bar A},\hat T_2^r\}$, with $T^{\bar A}$ being the 10 generators of the unbroken $SO(5)$ and $\hat T_2^r$ being the 5 generators of the coset $SO(6)/SO(5)$. For the convenience of later discussion about the SM gauge interactions, we further choose $T^{\bar A}=\{T_L^a,T_R^a,\hat T_1^i\}$, where $\{T_L^a,T_R^a\}$ belong to the subgroup $SO(4)\cong SU(2)_L\times SU(2)_R$ in $SO(5)$, while $\hat T_1^i$ are the generators of the coset $SO(5)/SO(4)$. The subscripts vary in the ranges ($a=1,2,3$), ($i=1,...,4$) and ($r=1,...,5$). 

The $SO(6)/SO(5)$ breaking gives 5 pNGBs $\vec{\pi}=(\pi_1,...,\pi_5)^T$, which can be used to construct the Goldstone matrix
\be
U(\vec{\pi})=e^{i\frac{\sqrt2}{f}\pi_r\hat T^r_2},
\ee
with $f$ being the Goldstone decay constant. The building blocks of CCWZ Lagrangian are the $d$ and $e$ symbols, which are defined by the Maurer-Cartan form as follows 
\be\label{de_definition}
U^\dagger iD_\mu U=d_\mu^r \hat T_2^r+e^{\bar A}_\mu T^{\bar A}\equiv d_\mu+e_\mu,
\ee
where gauge covariant derivative is
\be\label{A_definition}
D_\mu=\partial_\mu-igA_\mu\equiv\partial_\mu-igW_\mu^aT_L^a-ig' B_\mu T_R^3,
\ee
i.e. the SM gauge group $SU(2)_L\times U(1)_Y$ is embedded into the subgroup $SU(2)_L\times SU(2)_R\subset SO(5)$, where $Y=T_R^3$. The $\vec{\pi}$ as a \5 in $SO(5)$ can be decomposed under the SM gauge group as $\5\to\2_{1/2}\oplus\2_{-1/2}\oplus\1_0$, where $\2_{1/2}$ is the Higgs doublet
\be
H=\frac{1}{\sqrt{2}}\begin{pmatrix}\pi_2+i\pi_1\\ \pi_4-i\pi_3\end{pmatrix},
\ee
and $\2_{-1/2}$ is just the charge conjugate of $H$, while $\1_0$ is the real singlet $\pi_5$. The kinetic term of the pNGBs is constructed using the $d$ symbol, i.e. $\mL_{\rm kin}=(f^2/4)\,\tr\left[d_\mu d^{\mu}\right]$. To simplify the discussion, we adopt the unitary gauge by setting $\pi_{1,2,3}=0$ and redefining $\pi_{4,5}$ as~\cite{Gripaios:2009pe}
\be\label{h_redefinition}
\frac{h}{f}=\frac{\pi_4}{\sqrt{\pi_4^2+\pi_5^2}}\sin\frac{\sqrt{\pi_4^2+\pi_5^2}}{f},\quad\frac{\eta}{f}=\frac{\pi_5}{\sqrt{\pi_4^2+\pi_5^2}}\sin\frac{\sqrt{\pi_4^2+\pi_5^2}}{f}.
\ee
Then the Goldstone kinetic term becomes 
\begin{multline}\label{NGB_kinetic}
\mL_{\rm kin}=\frac{1}{2}\partial_\mu h\partial^\mu h+\frac{1}{2}\partial_\mu\eta\partial^\mu\eta \\+\frac12\frac{(h\partial_\mu h+\eta\partial_\mu\eta)^2}{f^2-h^2-\eta^2}+\frac{g^2}{8}h^2\left[\left(W_\mu^1\right)^2+\left(W_\mu^2\right)^2+\left(W_\mu^3-\frac{g'}{g}B_\mu\right)^2\right].
\end{multline}
After EW symmetry breaking (EWSB), $h$ gets the vacuum expectation value (VEV), and the $W$, $Z$ bosons gain their masses. The $T$-parameter is zero at tree-level because the custodial symmetry $SU(2)_V\subset SU(2)_L\times SU(2)_R$ is preserved in the EW vacuum.

Another import feature of the NMCHM is the existence of composite resonances. According to their spins, we can classify those resonances into the vector mesons (spin-1) and the fermionic top partners (spin-$1/2$). In the CCWZ framework, the composite objects form representations of the unbroken $SO(5)$. We consider the vector resonances in \1\0 and \5, and denote them as $\rho_\mu=\rho_\mu^{\bar A}T^{\bar A}$ and $a_\mu=a_\mu^r\hat T_2^r$ respectively. The Lagrangian is constructed using the $d$ and $e$ symbols
\be\label{spin-1}
\mL_\rho=-\frac{1}{4}\tr\left[\rho_{\mu\nu}\rho^{\mu\nu}\right]+\frac{M_{\rho}^2}{2g_{\rho}^2}\tr\left[(g_{\rho}\rho_\mu-e_\mu)^2\right]-\frac{1}{4}\tr[a_{\mu\nu}a^{\mu\nu}]+\frac{M_a^2}{2}\tr\left[a_\mu a^\mu\right],
\ee
where the strong sector coupling constant $g_\rho\gg g$, $g'$, and the field strengths read
\be
\rho_{\mu\nu}=\partial_\mu\rho_\nu-\partial_\nu\rho_\mu-ig_\rho [\rho_\mu,\rho_\nu],\quad a_{\mu\nu}=\nabla_\mu a_\nu-\nabla_\nu a_\mu,
\ee
where $\nabla_\mu=\partial_\mu-ie_\mu$ is the $SO(6)/SO(5)$ covariant derivative. \Eq{spin-1} is understood as a summation of resonances with the same quantum number but increasing masses, e.g.
\begin{equation*}
-\frac{1}{4}\tr\left[\rho_{\mu\nu}\rho^{\mu\nu}\right]+\frac{M_{\rho}^2}{2g_{\rho}^2}\tr\left[(g_{\rho}\rho_\mu-e_\mu)^2\right]\to
\sum_{n=1}^{N_\rho}-\frac{1}{4}\tr\left[\rho_{(n)\mu\nu}\rho_{(n)}^{\mu\nu}\right]+\frac{M_{\rho(n)}^2}{2g_{\rho(n)}^2}\tr\left[(g_{\rho}\rho_{(n)\mu}-e_\mu)^2\right],
\end{equation*}
and $M_{\rho(n+1)}>M_{\rho(n)}$. This short notation is also used in the Lagrangian the top partners (see the next subsection). 

The $\rho$- and $a$-resonances decompose to multiplets under the SM gauge group~\cite{Bian:2019kmg}
\be\label{vector_decomposition}
\left[\begin{array}{ccccccccccccc}\1\0&\to&\3_0&\oplus&\1_1&\oplus&\1_0&\oplus&\1_{-1}&\oplus&\2_{1/2}&\oplus &\2_{-1/2}\\ \rho^{\bar A}&\to&\rho_L&\oplus&\rho_R^+&\oplus&\rho_R^0&\oplus&\rho_R^-&\oplus&\rho_D&\oplus&\tilde\rho_D\end{array}\right];
\quad
\begin{bmatrix}
\5&\to&\2_{1/2}&\oplus &\2_{-1/2}&\oplus&\1_0\\ a^{r}&\to& a_D&\oplus&\tilde a_D&\oplus& a_S
\end{bmatrix},
\ee
where $\tilde\rho_D=i\sigma^2\rho_D^*$ is the charge conjugate of $\rho_D$, and similar for $\tilde a_D$. The expressions for this decomposition is in Appendix~\ref{app:CCWZ}. Those vector resonances can be produced via Drell-Yan process or vector boson fusion at the LHC, and decay to a pair of light bosons (SM bosons or $\eta$), or fermions (SM quarks or top partners). The 139 fb$^{-1}$ LHC data have constrained $M_\rho\gtrsim4$ TeV, provided the dominant branching ratio is the SM di-boson ($W^\pm Z$, $W^+W^-$, etc)~\cite{Aad:2019fbh,Aad:2020ddw}. The bounds are released if other decay channels are also considerable. For example, if the decay to a pair of top partners kinematically opens, then it dominates the branching ratios and the bound on $M_\rho$ is weakened to $\sim2.5$ TeV~\cite{Liu:2018hum}. The collider phenomenology of vector resonances in NMCHM can be found in Refs.~\cite{Bian:2019kmg,Banerjee:2017wmg,Franzosi:2016aoo,Niehoff:2016zso}.

\subsection{The fermion sector}

The boson sector is fixed by the coset $SO(6)/SO(5)$ thus is universal for all NMCHMs. However, the fermion sector is model-dependent. Partial compositeness mechanism says the fermions should be embedded in the incomplete representation of $SO(6)$ and mix with the strong fermionic operators linearly~\cite{Agashe:2004rs,Contino:2010rs}, but one has the freedom to choose different embeddings and build various models~\footnote{For recent progress in the direction of Higgs quadratic divergences cancellation we refer to Ref.~\cite{Csaki:2017jby,Guan:2019qux}.  }. As mentioned in the introduction, embeddings in \1\5 and lower representations are not easy to trigger a SFOEWPT, while in this article we propose a novel scenario in which $q_L$ and $t_R$ are both embedded in the high dimensional representation $\2\0'$.

There are three dimension-20 representations for $SO(6)$~\cite{slansky1981group}, while $\2\0'$ is the one obtained by $\6\otimes\6=\1\oplus\1\5\oplus\2\0'$, i.e. the traceless symmetric representation~\footnote{This representation has been considered in a couple of collider phenomenological studies~\cite{Serra:2015xfa,Banerjee:2017wmg}.}. To provide the correct hypercharge for the fermions, an additional $U(1)_X$ must be introduced and $Y=X+T_R^3$. To see the structure of the $\2\0'$, we list below the decomposition chain under $SO(6)\times U(1)_X\to SO(5)\times U(1)_X\to SO(4)\times U(1)_X\to SU(2)_L\times U(1)_Y$:
\bea\label{decomposition}
\2\0'_{2/3}&\to&\1\4_{2/3}\oplus\5_{2/3}\oplus\1_{2/3}\nn \\
&\to&\left(\9_{2/3}\oplus\4_{2/3}\oplus\1_{2/3}\right)\oplus\left(\4_{2/3}\oplus\1_{2/3}\right)\oplus\1_{2/3}\\
&\to&\left[\left(\3_{5/3}\oplus\3_{2/3}\oplus\3_{-1/3}\right)\oplus\left(\2_{7/6}\oplus\2_{1/6}\right)\oplus\1_{2/3}\right]\oplus\left[\left(\2_{7/6}\oplus\2_{1/6}\right)\oplus\1_{2/3}\right]\oplus\1_{2/3}.\nn
\eea
There are two $\2_{1/6}$ inside the $\2\0'$, coming from the \1\4 and \5 representations of $SO(5)$, respectively. Therefore, there are two ways to embed $q_L$, namely
\be
q_L^{\2\0'_A}=\frac{1}{2}\begin{pmatrix}0_{4\times4} & q_L^\4 & 0_{4\times1} \\
 (q_L^\4)^T & 0 & 0\\
 0_{1\times4}&0&0\end{pmatrix},
 \quad
 q_L^{\2\0'_B}=\begin{pmatrix}0_{4\times4}&0_{4\times1}&q_L^\4\\
 0_{1\times4}&0&0\\
 (q_L^\4)^T&0&0
 \end{pmatrix},
\ee
where $q_L^\4\equiv\left(ib_L,b_L,it_L,-t_L\right)^T$. The general embedding is the superposition of them
\be\label{qL20}
q_L^{\2\0'}=q_L^{\2\0'_A}e^{i\phi_L}\cos\theta_L +q_L^{\2\0'_B}\sin\theta_L.
\ee
On the other hand, there are three $\1_{2/3}$ in $\2\0'$, coming respectively from the \1\4, \5 and \1 of the $SO(5)$ subgroup and yielding three embeddings:
\be\begin{split}
t_R^{\2\0'_A}=&~\frac{1}{2\sqrt{5}}\begin{pmatrix}
 -\mathbb{I}_{4\times4}\,t_R & 0_{4\times2} \\
0_{2\times4} & 2\left(\mathbb{I}_{2\times2}+\sigma^3\right)t_R
\end{pmatrix},\quad
t_R^{\2\0'_B}=\frac{1}{\sqrt{2}}\begin{pmatrix}
 0_{4\times4} & 0_{4\times2} \\
 0_{2\times4} & \sigma^1\,t_R
\end{pmatrix},\\
t_R^{\2\0'_C}=&~\frac{1}{\sqrt{30}}\begin{pmatrix}
 -\mathbb{I}_{5\times5}\,t_R & 0_{5\times1} \\
 0_{1\times5} & 5\,t_R
\end{pmatrix},
\end{split}\ee
where $\sigma^a$ are the Pauli matrices. The general embedding of $t_R$ is then
\be\label{tR20}
t_R^{\2\0'}=e^{i\phi_{R1}}\cos\theta_{R1}t_R^{\2\0'_A}+e^{i\phi_{R2}}\sin\theta_{R1}\cos\theta_{R2}t_R^{\2\0'_B}+\sin\theta_{R1}\sin\theta_{R2}t_R^{\2\0'_C}.
\ee

According to the decomposition \Eq{decomposition}, we consider the top partners with $X=2/3$ and in \1, \5 and \1\4 representations of $SO(5)$. The Lagrangian of top partners is
\begin{multline}\label{spin-1/2}
\mL_\Psi=\tr\left[\bar\Psi_{\1\4}\left(i\slashed{\nabla}+g'\frac23\slashed{B}-M_{\1\4}\right)\Psi_{\1\4}\right]\\
+\bar\Psi_{\5}\left(i\slashed{\nabla}+g'\frac23\slashed{B}-M_\5\right)\Psi_\5+\bar\Psi_{\1}\left(i\slashed{\partial}+g'\frac23\slashed{B}-M_\1\right)\Psi_\1,
\end{multline}
where $\Psi_{\1\4}$ and $\Psi_\5$ are respectively $5\times5$ and $5\times1$ matrices, and
\be
\nabla_\mu\Psi_{\1\4}=\left(\partial_\mu-2i\,e_\mu^{\bar A}t^{\bar A}\right)\Psi_{\1\4},\quad
\nabla_\mu\Psi_\5=\left(\partial_\mu-i\,e_\mu^{\bar A}t^{\bar A}\right)\Psi_\5,
\ee
and $[t^{\bar A}]_{rs}\equiv [T^{\bar A}]_{rs}$ with $(r,s=1,...,5)$. The factor 2 in the covariant derivative of $\Psi_{\1\4}$ is due to its symmetric structure. The top partners interact with the vector resonances strongly,
\be
\mL_{\rho\Psi}=c_{\1\4}\,\tr\left[\bar\Psi_{\1\4}\gamma^\mu t^{\bar A}\Psi_{\1\4}\right](g_{\rho}\rho^{\bar A}_\mu-e^{\bar A}_\mu)+
c_\5\bar\Psi_{\5}\gamma^\mu t^{\bar A}\Psi_{\5}(g_{\rho}\rho^{\bar A}_\mu-e^{\bar A}_\mu)+\cdots,
\ee
where $c_{\1\4,\5}$ are $\mO(1)$ numbers. Those vertices imply the vector resonances can decay to a pair of top partners (if kinematically allowed). Due to the large coupling $g_\rho$, once opened those channels will dominate branching ratio quickly~\cite{Liu:2018hum}. The interactions between the SM quarks and top partners are connected by the Goldstone matrix,
\be\begin{split}\label{Yukawa}
\mL_{q\Psi}=&~ y_L^{\1\4}f\left(\bar q_L^{\2\0'}\right)_{IJ}U_{Ir}U_{Js}\Psi_{\1\4}^{rs}+y_R^{\1\4}f\left(\bar t_R^{\2\0'}\right)_{IJ}U_{Ir}U_{Js}\Psi_{\1\4}^{rs}+\hc\\
&+y_L^{\5}f\left(\bar q_L^{\2\0'}\right)_{IJ}U_{Ir}\Sigma_J\Psi_{\5}^{r}+y_R^{\5}f\left(\bar t_R^{\2\0'}\right)_{IJ}U_{Ir}\Sigma_J\Psi_{\5}^{r}+\hc\\
&+y_L^{\1}f\left(\bar q_L^{\2\0'}\right)_{IJ}\Sigma_I\Sigma_J\Psi_\1+y_R^{\1}f\left(\bar t_R^{\2\0'}\right)_{IJ}\Sigma_I\Sigma_J\Psi_\1+\hc~,
\end{split}\ee
where $y_{L,R}$ are mixing parameters, and the indices $(I,J=1,...,6)$. The Goldstone vector is $\Sigma=U\Sigma_0$, where $\Sigma_0=(0,0,0,0,0,1)^T$ is the $SO(5)$-preserving vacuum state.

The top partner decompositions under the SM gauge group are
\be\label{fermion_decomposition_2}
\left[\begin{array}{ccccccccccccc}
\1\4_{2/3}&\to &\3_{5/3}&\oplus&\3_{2/3}&\oplus&\3_{-1/3}&\oplus&\2_{7/6}&\oplus&\2_{1/6}&\oplus&\1_{2/3}\\
 \Psi_{\1\4}&\to& K&\oplus& N&\oplus &Y&\oplus &J_X&\oplus &J_Q&\oplus& T'
\end{array}\right],
\ee
and
\be\label{fermion_decomposition}
\begin{bmatrix}
\5_{2/3}&\to&\2_{7/6}&\oplus &\2_{1/6}&\oplus&\1_{2/3}\\ \Psi_\5&\to& Q_X&\oplus& Q&\oplus&\widetilde T\end{bmatrix},
\ee
from which we get a set of vector-like quarks (VLQ) with electric charges varying from $8/3$ to $-4/3$ with a step size of 1. Again, the full expressions of the decomposition are given in Appendix~\ref{app:CCWZ}. While the VLQs with exotic charge $8/5$, $5/3$ or $-4/3$ are already in their mass eigenstates, the ones with charge $2/3$ and $-1/3$ mix with the SM third generation quarks after EWSB, and mass eigenstates should be extracted by diagonalizing the mass matrices. The SM bottom quark remains massless after such a diagonalization, because we don't include $b_R$ yet in \Eq{Yukawa}. On the other hand, $b_L$ mixes with the VLQs with charge $-1/3$. For example,
\be\label{bL_mix}
\mL_{q\Psi}\supset-\frac{1}{\sqrt{2}}
h y_L^{\1\4} \left(\frac{\eta e^{-i \phi_L}\cos\theta_L}{f+\sqrt{f^2-h^2-\eta ^2}}+\sin\theta_L\right)(\bar b_LN_{-1/3}+\bar b_LY_{-1/3}),
\ee
implying the $b_L$-$N_{-1/3}$ and $b_L$-$Y_{-1/3}$ mixing after EWSB, where $N_{-1/3}$ and $Y_{-1/3}$ denote the charge $-1/3$ component of the $N$ and $Y$ triplet, respectively. Such a mixing changes the coupling between left-handed fermion and the $Z$ boson, which is
\be
\frac{g}{c_W}\left(T_L^3-s_W^2Q\right),
\ee
for a fermion with third-component weak isospin $T_L^3$ and charge $Q$. As $T_L^3(b_L)=-1/2$, $T_L^3(N_{-1/3})=-1$ and $T_L^3(Y_{-1/3})=0$, the mixing in \Eq{bL_mix} gives a large correction to the $Zb_L\bar b_L$ coupling, which is unacceptable because this vertex has been measured at the LEP at a very high accuracy~\cite{ALEPH:2005ab,Gori:2015nqa}. One proper way to avoid this problem is to choose $\theta_L=0$ in the $q_L^{\2\0'}$ embedding of \Eq{qL20}, and require $\ave{\eta}=0$ at zero temperature. The mixing terms in \Eq{bL_mix} then vanish. That means we use $q_L^{\2\0'}\equiv q_L^{\2\0'_A}$ in the model from now on. The mixing between $b_L$ and the charge $-1/3$ top partners from $SO(4)$ bi-doublets (such as $Q$ or $J_Q$) is safe because the $P_{LR}$ symmetry protects the $Zb_L\bar b_L$ vertex~\cite{Agashe:2006at}.

After the embedding of $q_L$ is fixed, different choices of the $t_R$ embedding (i.e. the parameters $\theta_{R1}$, $\theta_{R2}$, etc in \Eq{tR20}) give different form factors in \Eq{tLR}. For example,
\be\label{PiLR}\begin{split}
t_R^{\2\0'_A}:&~\Pi_{LR}^t=-\frac{1}{2\sqrt{5}}\frac{h\eta}{f^2}\left(\frac{3M_1^t}{2}-M_2^t\frac{h^2-4\eta^2}{f^2}\right),\\
t_R^{\2\0'_B}:&~\Pi_{LR}^t=-\frac{1}{2\sqrt{2}}\frac{h}{f}\sqrt{1-\frac{h^2+\eta^2}{f^2}}\left(M_1^t+4M_2^t\frac{\eta^2}{f^2}\right),\\
t_R^{\2\0'_C}:&~\Pi_{LR}^t=\frac{1}{\sqrt{30}}\frac{h\eta}{f^2}\left[M_1^t-M_2^t\left(5-6\frac{h^2+\eta^2}{f^2}\right)\right].
\end{split}\ee
Since $\Big|\Pi_{LR}^t\big|_{p^2=0}\Big|$ is the top quark mass, from \Eq{PiLR} one finds that only the $t_R^{\2\0'_B}$ embedding gives a massive top when $\ave{\eta}=0$. Since $\ave{\eta}=0$ is needed for a SM-like $Zb_L\bar b_L$, we conclude that the $t_R^{\2\0'}$ embedding must have a non-zero $t_R^{\2\0'_B}$ component, i.e. $\sin\theta_{R1}\cos\theta_{R2}\neq0$ in \Eq{tR20}. For simplicity, we will only deal with $t_R^{\2\0'_B}$ in the rest of this article and this can be understood as we assign an odd $\Z_2$ number for $\eta$ in the third generation quark embeddings. In summary, based on the $Zb_L\bar b_L$ vertex and the top mass constraints, hereafter we will consider the combination $q_L^{\2\0'_A}+t_R^{\2\0'_B}$ as the $\2\0'+\2\0'$ NMCHM.

The top partners can be produced at the LHC either in pair via QCD or singly via EW fusion, and finally decay to a SM fermion plus boson(s) (e.g $bW^+$, $tW^+$, $t\eta$, etc). Searches for the pair production VLQs with charge $5/3$ or $2/3$ have set limits of $M_{\5},~M_{\1\4}\gtrsim1.3$ TeV at the LHC with an integrated luminosity of $\approx36{\rm ~fb}^{-1}$~\cite{Sirunyan:2018yun,Aaboud:2018pii}, while the bounds from single production are typically weaker~\cite{Aaboud:2018xpj,Sirunyan:2018ncp}. $\Psi_\1$ mainly decays to $t\eta$ via the term $y_R^{\1}\bar t_R\Psi_\1\eta\subset\mL_{q\Psi}$, and the constraints can be as weak as $M_\1\lesssim1$ TeV~\cite{Cacciapaglia:2019zmj}. About the collider phenomenology of the VLQs in the CHMs, see Refs.~\cite{Serra:2015xfa,Banerjee:2017qod,Banerjee:2017wmg,Franzosi:2016aoo,Niehoff:2016zso,Bizot:2018tds,Xie:2019gya,Cacciapaglia:2019zmj} for the charge $5/3$ and $2/3$ ones and Ref.~\cite{Matsedonskyi:2014lla} for the charge $8/3$ one (coming from the $K$ triplet).

\subsection{Detailed expressions for formulae}\label{app:CCWZ}

First we present the $SO(6)$ generators~\cite{Niehoff:2016zso}:
\be\begin{split}
[T^{a}_L]_{IJ}=&-\frac{i}{2}\left[\frac{1}{2}\epsilon^{abc}(\delta_{bI}\delta_{cJ}-\delta_{bJ}\delta_{cI})+(\delta_{aI}\delta_{4J}-\delta_{aJ}\delta_{4I})\right],\\
[T^{a}_R]_{IJ}=&-\frac{i}{2}\left[\frac{1}{2}\epsilon^{abc}(\delta_{bI}\delta_{cJ}-\delta_{bJ}\delta_{cI})-(\delta_{aI}\delta_{4J}-\delta_{aJ}\delta_{4I})\right],\\
[\hat T^{i}_1]_{IJ}=&-\frac{i}{\sqrt2}(\delta_{iI}\delta_{5J}-\delta_{iJ}\delta_{5I}),\\
[\hat T^{r}_2]_{IJ}=&-\frac{i}{\sqrt2}(\delta_{rI}\delta_{6J}-\delta_{rJ}\delta_{6I}),
\end{split}\ee
where the indices ranges are ($a=1,2,3$), ($i=1,\cdots,4$), ($r=1,\cdots,5$) and ($I,J=1,\cdots,6$). This definition yields a normalization of $\tr[T^AT^B]=\delta^{AB}$.

Next we give the explicit expressions for the $d$ and $e$ symbols defined in \Eq{de_definition} in unitary gauge~\cite{Bian:2019kmg}. The $d$ symbols are
\be\begin{split}
d^1_\mu=&~\frac{gW_\mu^1}{\sqrt{2}}\frac{h}{f},\quad d^2_\mu=\frac{gW_\mu^2}{\sqrt{2}}\frac{h}{f},\quad 
d^3_\mu=\frac{gW_\mu^3-g'B_\mu}{\sqrt{2}}\frac{h}{f},\\
d^4_\mu=&~\frac{\sqrt{2}}{f}\frac{1}{h^2+\eta^2}\left[\eta\left(h\partial_\mu\eta-\eta\partial_\mu h\right)-\frac{h\left(h\partial_\mu h+\eta\partial_\mu\eta\right)}{\sqrt{1-(h^2+\eta^2)/f^2}}\right],\\
d^5_\mu=&~\frac{\sqrt{2}}{f}\frac{1}{h^2+\eta^2}\left[h\left(\eta\partial_\mu h-h\partial_\mu \eta\right)-\frac{\eta\left(h\partial_\mu h+\eta\partial_\mu\eta\right)}{\sqrt{1-(h^2+\eta^2)/f^2}}\right];
\end{split}\ee
while the $e$ symbols are decomposed to $e^{\bar A}_{\mu}=\{e_{L\mu}^{a},e_{R\mu}^{a},e_{1\mu}^{i}\}$ under the $SO(4)$ subgroup, yielding
\be\begin{split}
e_{L\mu}^{1}=&~gW^1_\mu-\frac{1}{2}gW^1_\mu\frac{h^2}{f^2}\left(\frac{1}{1+\sqrt{1-(h^2+\eta^2)/f^2}}\right),\\
e_{L\mu}^{2}=&~gW^2_\mu-\frac{1}{2}gW^2_\mu\frac{h^2}{f^2}\left(\frac{1}{1+\sqrt{1-(h^2+\eta^2)/f^2}}\right),\\
e_{L\mu}^{3}=&~gW^3_\mu-\frac{1}{2}\left(gW^3_\mu-g'B_\mu\right)\frac{h^2}{f^2}\left(\frac{1}{1+\sqrt{1-(h^2+\eta^2)/f^2}}\right);
\end{split}\ee
and
\be\begin{split}
e_{R\mu}^{1}=&~\frac{1}{2}gW^1_\mu\frac{h^2}{f^2}\left(\frac{1}{1+\sqrt{1-(h^2+\eta^2)/f^2}}\right),\\
e_{R\mu}^{2}=&~\frac{1}{2}gW^2_\mu\frac{h^2}{f^2}\left(\frac{1}{1+\sqrt{1-(h^2+\eta^2)/f^2}}\right),\\
e_{R\mu}^{3}=&~g'B_\mu+\frac{1}{2}\left(gW^3_\mu-g'B_\mu\right)\frac{h^2}{f^2}\left(\frac{1}{1+\sqrt{1-(h^2+\eta^2)/f^2}}\right);
\end{split}\ee
and
\be\begin{split}
e_{1\mu}^{1}=&~-\frac{1}{\sqrt{2}}gW^1_\mu\frac{h\eta}{f^2}\left(\frac{1}{1+\sqrt{1-(h^2+\eta^2)/f^2}}\right),\\
e_{1\mu}^{2}=&~-\frac{1}{\sqrt{2}}gW^2_\mu\frac{h\eta}{f^2}\left(\frac{1}{1+\sqrt{1-(h^2+\eta^2)/f^2}}\right),\\
e_{1\mu}^{3}=&~-\frac{1}{\sqrt{2}}\left(gW^3_\mu-g'B_\mu\right)\frac{h\eta}{f^2}\left(\frac{1}{1+\sqrt{1-(h^2+\eta^2)/f^2}}\right),\\
e_{1\mu}^{4}=&~\sqrt{2}\frac{\eta\partial_\mu h-h\partial_\mu\eta}{f^2}\left(\frac{1}{1+\sqrt{1-(h^2+\eta^2)/f^2}}\right).
\end{split}\ee

Now we turn to the resonances. The full expressions of the vector resonances decomposition in \Eq{vector_decomposition} are~\cite{Bian:2019kmg} 
\be\begin{split}
\rho_{L\mu}^{\pm}=&~\frac{\rho_{L\mu}^1\mp i\rho_{L\mu}^2}{\sqrt{2}},\quad \rho_{L\mu}^{0}=\rho_{L\mu}^3;\quad
\rho_{R\mu}^{\pm}=\frac{\rho_{R\mu}^1\mp i\rho_{R\mu}^2}{\sqrt{2}},\quad \rho_{R\mu}^{0}=\rho_{R\mu}^3;\\
\rho_{D\mu}=&~\begin{pmatrix}\rho_{D\mu}^+\\ \rho_{D\mu}^0\end{pmatrix}=\frac{1}{\sqrt{2}}\begin{pmatrix}\rho_{1\mu}^2+i\rho_{1\mu}^1\\ \rho_{1\mu}^4-i\rho_{1\mu}^3\end{pmatrix},\\
a_{D\mu}=&~\begin{pmatrix}a_{D\mu}^+\\ a_{D\mu}^0\end{pmatrix}=\frac{1}{\sqrt{2}}\begin{pmatrix}a_{\mu}^2+ia_{\mu}^1\\ a_{\mu}^4-ia_{\mu}^3\end{pmatrix},\quad a_{S\mu}=a_\mu^5.
\end{split}\ee
After the decomposition, we have 4 singly charged and 7 real neutral vector resonances, in total 15 degrees of freedom.

Finally we give the details of the top partner decompositions listed in Eqs.~(\ref{fermion_decomposition_2}) and (\ref{fermion_decomposition}). As the \1\4 of the $SO(5)$, $\Psi_{\1\4}$ can first decompose to 3 multiplets under the $SO(4)$ subgroup, i.e.
\be
\Psi_{\1\4}=\begin{pmatrix}K_{(\3,\3)}&0_{4\times1}\\ 0_{1\times4}&0\end{pmatrix}+\frac{1}{\sqrt{2}}\begin{pmatrix}0_{4\times4}&J_{(\2,\2)}\\ J_{(\2,\2)}^T&0\end{pmatrix}+\frac{1}{2\sqrt{5}}\begin{pmatrix}-\mathbb{I}_{4\times4}T'&0_{4\times1}\\ 0_{1\times4}&4T'\end{pmatrix},
\ee
where $K_{(\3,\3)}$, $J_{(\2,\2)}$ and $T'$ are in $(\3,\3)$, $(\2,\2)$ and $(\1,\1)$ of $SO(4)$, respectively. Under the SM gauge group, $K_{(\3,\3)}$ further decompose to three $SU(2)_L$ triplets with hypercharges $5/3$, $2/3$ and $-1/3$, while $J_{(\2,\2)}$ decomposes to two $SU(2)_L$ doublets with hypercharges $7/6$ and $1/6$. Explicitly, they are
\be\begin{split}
&K_{(\3,\3)}=\\
&\begin{tiny}\begin{pmatrix}\frac{1}{2} (K_{8/3}+N_{2/3}+Y_{-4/3}) & \frac{1}{2} i
   (K_{8/3}-Y_{-4/3}) & \frac{-K_{5/3}+N_{5/3}-N_{-1/3}+Y_{-1/3}}{2 \sqrt{2}} &
   \frac{i (K_{5/3}+N_{5/3}+N_{-1/3}+Y_{-1/3})}{2 \sqrt{2}} \\
 \frac{1}{2} i (K_{8/3}-Y_{-4/3}) & \frac{1}{2}
   (-K_{8/3}+N_{2/3}-Y_{-4/3}) & -\frac{i
   (K_{5/3}-N_{5/3}-N_{-1/3}+Y_{-1/3})}{2 \sqrt{2}} &
   \frac{-K_{5/3}-N_{5/3}+N_{-1/3}+Y_{-1/3}}{2 \sqrt{2}} \\
 \frac{-K_{5/3}+N_{5/3}-N_{-1/3}+Y_{-1/3}}{2 \sqrt{2}} & -\frac{i
   (K_{5/3}-N_{5/3}-N_{-1/3}+Y_{-1/3})}{2 \sqrt{2}} & \frac{1}{2}
   (K_{2/3}-N_{2/3}+Y_{2/3}) & -\frac{1}{2} i
   (K_{2/3}-Y_{2/3}) \\
 \frac{i (K_{5/3}+N_{5/3}+N_{-1/3}+Y_{-1/3})}{2 \sqrt{2}} &
   \frac{-K_{5/3}-N_{5/3}+N_{-1/3}+Y_{-1/3}}{2 \sqrt{2}} & -\frac{1}{2} i
   (K_{2/3}-Y_{2/3}) & \frac{1}{2} (-K_{2/3}-N_{2/3}-Y_{2/3}) \end{pmatrix}\end{tiny},
\end{split}\ee
where
\be
K=\begin{pmatrix}K_{8/3}\\ K_{5/3}\\ K_{2/3}\end{pmatrix},\quad
N=\begin{pmatrix}N_{5/3}\\ N_{2/3}\\ N_{-1/3}\end{pmatrix},\quad
Y=\begin{pmatrix}Y_{2/3}\\ Y_{-1/3}\\ Y_{-4/3}\end{pmatrix},
\ee
are the multiplets with SM quantum number $\3_{5/3}$, $\3_{2/3}$ and $\3_{-1/3}$ respectively; and
\be
J_{(\2,\2)}=\frac{1}{\sqrt{2}}\left( i J_{-1/3}-i J_{5/3}, J_{5/3}+J_{-1/3}, i J_{2/3A}+iJ_{2/3B} , J_{2/3B}-J_{2/3A}\right)^T,
\ee
where
\be
J_X=\begin{pmatrix}J_{5/3}\\ J_{2/3B}\end{pmatrix},\quad J_Q=\begin{pmatrix}J_{2/3A}\\ J_{-1/3}\end{pmatrix}
\ee
are the multiplets with SM quantum number $\2_{7/6}$ and $\2_{1/6}$ respectively. Another top partner $\Psi_\5$ is written as
\be
\Psi_\5 = \frac{1}{\sqrt{2}}\left(
i B - i X_{5/3},
 B + X_{5/3},
i T + i X_{2/3},
 -T + X_{2/3},
 \widetilde T \right)^T,
\ee
in which two $SU(2)_L\times U(1)_Y$ doublets
\be
Q_X=\begin{pmatrix}X_{5/3}\\ X_{2/3}\end{pmatrix}_{7/6},\quad Q=\begin{pmatrix}T\\ B\end{pmatrix}_{1/6},
\ee
and one singlet $\widetilde T$ with hyper charge $2/3$ are present. 

\bibliographystyle{JHEP-2-2.bst}
\bibliography{references}

\end{document}